\documentclass[12pt]{article}
\usepackage{epsfig}

\def\an#1#2#3{{Ann. Rev. Nucl. Part. Sci. {\bf #1} #2 (#3)}}

\def\ijm#1#2#3{{Int. J. Mod. Phys. {\bf A~#1} #2 (#3)}}

\def\nc#1#2#3{{Nuovo Cimento {\bf #1~A} #2 (#3)}}
\def\npb#1#2#3{{Nucl. Phys. {\bf B~#1} #2 (#3)}}
\def\plb#1#2#3{{Phys. Lett. {\bf B~#1} #2 (#3)}}
\def\prd#1#2#3{{Phys. Rev. {\bf D~#1} #2 (#3)}}
\def\prl#1#2#3{{Phys. Rev. Lett. {\bf #1} #2 (#3)}}

\def\zp#1#2#3{{Z. Phys. {\bf C~#1} #2 (#3)}}
\def\etal{{\it et al.}}

\def\ri{\rightarrow}

\def\gsim{\ \rlap{\raise 2pt \hbox{$>$}}{\lower 2pt \hbox{$\sim$}}\ }
\def\lsim{\ \rlap{\raise 2pt \hbox{$<$}}{\lower 2pt \hbox{$\sim$}}\ }
\begin{document}

\sloppy
\vskip -2.2truecm
{\hfill UMHEP-410}

\bigskip\bigskip\medskip

\centerline{\Large {\bf Reanalysis of the decay \mbox{\boldmath 
$K_L \rightarrow \pi^0 e^+ e^-$}}}

\medskip
\bigskip\bigskip\medskip

\centerline{\large {John F. Donoghue and Fabrizio Gabbiani}}
\medskip

\centerline{\large{Department of Physics and Astronomy}}
\smallskip

\centerline{\large{University of Massachusetts, Amherst, Massachusetts
01003}}

\vskip 0.9 truecm

\begin{abstract}

The decay $K_L \rightarrow \pi^0 e^+ e^-$ is being searched for as a signal
of direct $\Delta S = 1$ CP violation. We provide a thorough updating of
the analysis of the three components of the decay: 1) Direct CP violation, 2)
CP violation through the mass matrix and 3) CP-conserving (two-photon)
contributions. First the chiral calculation of the $K_S \rightarrow \pi^0
e^+ e^-$ rate, due to Ecker, Pich and de Rafael (EPR), is updated to include
recent results on the nonleptonic amplitude. Then we systematically explore
the uncertainties in this method. These appear to be so large that they will
obscure the direct CP violation unless it is possible to measure the $K_S
\rightarrow \pi^0 e^+ e^-$ rate. The CP-conserving amplitude remains
somewhat uncertain, but present indications are such that there may be a
sizable CP-violating asymmetry in the $e^+, e^-$ energies from the
interference of the CP-conserving and CP-violating amplitudes and this may
potentially be useful in determining whether direct CP violation is present.

\end{abstract}

\section{Introduction}

One of the goals of the next generation of rare kaon decay experiments
is to attempt to observe CP violation in the decay $K_L \rightarrow
\pi^0 e^+ e^-$. This reaction is special because we expect that
direct CP violation (as opposed to the ``mass matrix" CP violation
already observed in the parameter $\epsilon$) may be the dominant
component of the amplitude. This is in contrast with $K_L \rightarrow
\pi \pi$, where the direct effect is at most a few parts in a thousand
of $\epsilon$. Direct CP violation distinguishes the Standard Model
from ``superweak"-type models \cite{SW}. Moreover, the magnitude of
the direct CP violation for this reaction is a precise prediction
of the Standard Model,
with very little hadronic uncertainty. In this article, we will
update the analysis of the reaction $K_L \rightarrow \pi^0 e^+ e^-$,
attempting to understand if we can be certain that an experimental
measurement is in fact a signal of direct CP violation.

One difficulty is that there are three possible components to the decay
amplitude: 1) A CP-conserving process which proceeds through two-photon
exchanges, 2) a mass matrix CP-violating effect proportional to the known
parameter $\epsilon$, and 3) the direct $\Delta S = 1$ CP-violating effect
which we would like to observe. The existence of the first of these
indicates that simply observing the total decay rate is not sufficient to
unambiguously indicate the existence of CP violation. We need to either
observe a truly CP-odd decay asymmetry, or else be confident on the basis
of a theoretical calculation that the CP-conserving effect is safely smaller
than the experimental signal. Unfortunately, the predictions in the literature
for each of the components listed above exhibits a range of values,
including some estimates where all three are similar in magnitude.
However, the quality of the theoretical treatment can improve with time,
effort and further experimental input. We will try to assess the present and
future uncertainties in the theoretical analysis.

There remain significant experimental difficulties before it is possible to
mount a search sensitive to a branching ratio of a few times $10^{-12}$.
We will assume that such a sensitivity is reached. At the same time, it is
reasonable to assume that we will have improved experimental information
on the related rate $K_L \rightarrow \pi^0 \gamma \gamma$, and that
theoretical methods have provided a consistent phenomenology of this
reaction.

%\vfill\eject.

\noindent The KM parameters will be somewhat more fully constrained in
the future, but hadronic matrix element uncertainties will prevent a precise
determination of the parameters relevant for CP violation, at least until B
meson CP violation has been extensively explored. With these expectations
as our framework, will we be able to prove that the future experimental
observation indicates the presence of direct CP violation?

Our analysis indicates that one will not be able to prove the existence of
direct CP violation from the branching ratio for
$K_L \rightarrow \pi^0 e^+ e^-$ unless
the decay rate for the related decay $K_S \rightarrow \pi^0 e^+ e^-$ is also
observed experimentally. This is yet more difficult than measuring the
$K_L$ decay, and poses a problem for the program of finding direct CP
violation. It is possible but not certain, that the electron charge asymmetry
can resolve this issue and,
when combined with the rate, signal direct CP violation.

\section{Overview of the analysis}

There is an extensive analysis associated with each of the three components
of the decay amplitude which were listed in the introduction. We will
devote separate sections of the paper to each of the major issues. The
purpose of the present section is to highlight the main issues which are to be
discussed more fully later, and to indicate how they fit together in an overall
description of the decay process.

In a way, the direct CP component is the simplest. The uncertainties are
only in the basic parameters of the Standard Model, i.e., the mass of the top
quark and the KM parameters. The relevant hadronic matrix element is
reliably known. Unfortunately the extraction of KM elements has
significant uncertainties, so that only a range of possible values can be
given. This range corresponds to $K_L \rightarrow \pi^0 e^+ e^-$
branching ratios of a few times $10^{-12}$. We discuss this range in
Sec. III.

The contribution of mass-matrix CP violation is more uncertain. The
rate due to this source is given by the parameter $\epsilon$ times the
rate for $K_S \rightarrow \pi^0 e^+ e^-$, so that the issue is the
prediction of the CP-conserving $K_S$ partial rate. Here the primary
tool is chiral perturbation theory, with the pioneering treatment
given by Ecker, Pich and de Rafael (EPR) \cite{EPR1,EPR2}. In Sec.
IV, we update their analysis, under essentially the same assumptions.
The main new ingredient is the inclusion of the results of the one
loop analysis of nonleptonic decays, which decreases the overall
strength of the weak $K \rightarrow \pi$ transition. This yields a
change in the weak counterterms and a decrease of the rate. However,
more important is an assessment of the uncertainties of such a
calculation, which we describe in Sec. V. Any such calculation has a
range of uncertainties, most of which we are able to estimate based on
past experience with chiral calculations. We systematically discuss
these. Unfortunately we find that one issue in particular has a
devastating sensitivity on this mode. In their analysis, EPR made an
assumption which lies outside of chiral perturbation theory, that a
certain weak counterterm satisfies $w_2 = 4 L_9$ where $L_9$ is a
known coefficient in the QCD effective chiral Lagrangian. This
assumption has no rigor, and the decay rate is very sensitive to the
deviation $w_2 - 4 L_9$. For any reasonable value of direct CP
violation, there is an equally reasonable value of $w_2$ which can
reproduce the corresponding $K_L \rightarrow \pi^0 e^+ e^-$ decay
rate. Given a measurement, we will then be intrinsically unable to
decide if it is evidence of a nonzero value of direct CP violation or
merely measures a value for $w_2$. It is this which indicates a need
to measure the rate $K_S \rightarrow \pi^0 e^+ e^-$.

The third component is the CP-conserving amplitude which proceeds
through the two-photon intermediate state $K_L \rightarrow \pi^0 \gamma
\gamma , \gamma \gamma \rightarrow e^+ e^-$, described in Sec. VI.
Here we must first understand the process $K_L \rightarrow \pi^0 \gamma
\gamma$. This has been calculated in chiral perturbation theory at one loop
order and has been measured experimentally. While the shape agrees with
the chiral calculation, the theoretical rate misses by a factor of three. This
has prompted some reanalyses of the theory of $K_L \rightarrow \pi^0
\gamma \gamma$, which we will take account of. However, the field has
not reached a full conclusion on this mode, and it is clear that in the future
the experimental and phenomenological status of this reaction will
undoubtedly improve. We study how possible resolutions of these analyses
will influence the $K_L \rightarrow \pi^0 \gamma \gamma$ decay rate.
Ultimately this component should be satisfactorily understood.

The ultimate problem is then our inability to distinguish, in a measurement
of the $K_L \rightarrow \pi^0 e^+ e^-$ decay rate, the direct CP violation
from the mass matrix effect. It is possible that the electron energy
asymmetry may allow us to make this separation. The electron asymmetry
comes from the interference of the CP-conserving two-photon process (even
under the interchange of $e^+ e^-$) and the CP-violating one-photon
process (odd under the $e^+ e^-$ interchange). For many values of the
presently favored parameter range, this asymmetry is very large i.e., of
order 50\%. In this case its measurement is not far more difficult than a
good measurement of the rate. If we in fact are able to reach an
understanding of the two-photon process, through future phenomenology
and experiments on $K_L \rightarrow \pi^0 \gamma \gamma$, then the
asymmetry depends most critically on the CP-violating amplitude. If there
is no direct CP violation, there is then a correlation between the $K_L
\rightarrow \pi^0 e^+ e^-$ decay rate and the electron asymmetry,
parameterized by the unknown coefficient $w_2$. As we detail in Sec.
VII, the presence of direct CP violation would upset this correlation, and in
many cases would lead to a drastically different relative size of the
asymmetry vs. decay rate, often even changing the sign of the asymmetry.
Thus the asymmetry may be used to signal direct CP violation.
Unfortunately this method is not foolproof. There exists combinations of
values of $w_2$ and KM angles for whom the distinction between direct
and mass matrix CP violation are not so great and will be muddied by the
inherent uncertainties in the theory. In Sec. VIII we explore the use of
$K_L - K_S$ interference to sort out the direct CP-violating amplitude.

Overall, our reanalysis indicates that the demands on the experimental
exploration of this reaction are quite severe. The simple observation of a
few events will not be sufficient to indicate direct CP violation. The
measurement of an electron asymmetry requires more events, and may or
may not resolve the issue. Only the simultaneous measurement of $K_S
\rightarrow \pi^0 e^+ e^-$ allows a convincing proof of the existence of
direct CP violation.

\section{Direct CP violation}

Direct $\Delta S = 1$ CP violation is manifested in the ``penguin" reactions
pictured in Fig. 1. The QCD short distance corrections to this mode have
been thoroughly analyzed to next-to-leading order by Buras \etal\ \cite{B},
and we will use their results. The primary weak operator responsible for the
transition have the form

\begin{equation}
{\cal H}^{\phantom{\dagger}}_W = {G_F \over \sqrt{2}}
\left[ C_{7V} (\mu) Q_{7V} + C_{7A} Q_{7A} \right],
\end{equation}

\noindent where

\begin{eqnarray}
Q_{7V} & = & ( \bar{s} d)_{V-A} (\bar{e} e)_V, \nonumber \\
Q_{7A} & = & ( \bar{s} d)_{V-A} (\bar{e} e)_A.
\end{eqnarray}

\noindent The dominant contribution to the imaginary part of the coefficient
$C_{7i}$ comes from the top quark, so that this is truly a short distance
process. The coefficients have a CP-violating component

\begin{equation}
{\rm Im} \; C_{7i} = -{\rm Im} \left( {V^{\phantom{*}}_{td} V^*_{ts} \over
V^{\phantom{*}}_{ud} V^*_{us}}
\right) y^{\phantom{\dagger}}_{7i},
\end{equation}

\noindent with the results of Ref. \cite{B} yielding

\begin{equation}
{y_{7V} \over \alpha} = \left\{\begin{array}{l}
0.708 \quad m_t = 150, \\
0.743 \quad m_t = 175, \\
0.775 \quad m_t = 200, \end{array}\right. \nonumber \\
\qquad {y_{7A} \over \alpha} = \left\{\begin{array}{l}
-0.579 \quad m_t = 150, \\
-0.736 \quad m_t = 175, \\
-0.905 \quad m_t = 200, \end{array}\right.
\end{equation}

\noindent with very little dependence on $\Lambda_{\overline{MS}}$ (the
above is for $\Lambda_{\overline{MS}}$ = 0.3 GeV) and a negligible
dependence on the low energy scale $\mu$.

\begin{figure}[!htbp]
\centering
\leavevmode
\centerline{
\epsfbox{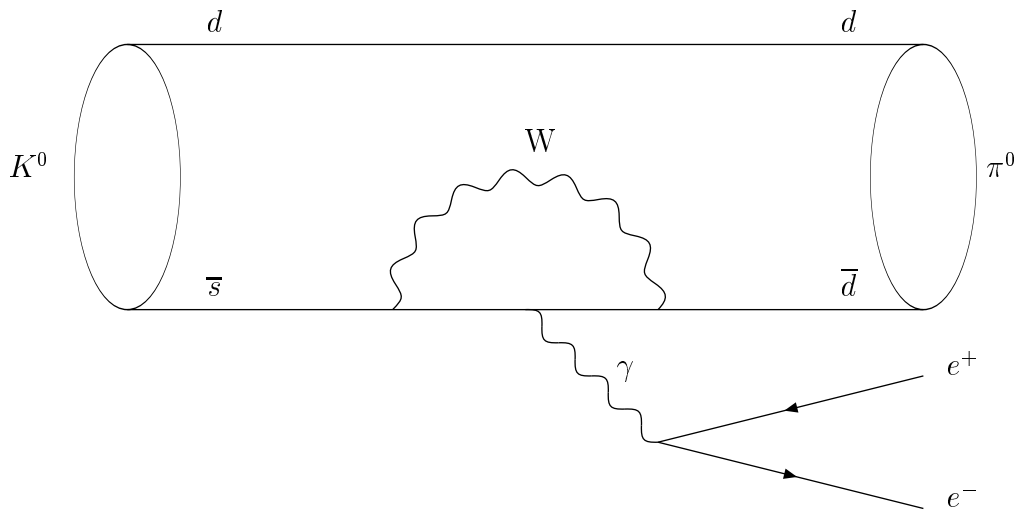}}
\caption{``Penguin'' diagram inducing $\Delta$S = 1 CP
violation.}
\end{figure}

The matrix element involved is well known via isospin symmetry from the
charged current kaon decay, i.e.,

\begin{equation}
\langle \pi^0 (p^{\prime}) \vert \bar{d} \gamma_{\mu} s \vert K^0 (p) \rangle =
{f_+ (q^2) \over \sqrt{2}} (p + p)_{\mu} + {f_- (q^2) \over \sqrt{2}} (p -
p^{\prime})_{\mu},
\end{equation}

\noindent with

\begin{eqnarray}
f_+ (q^2) & = & 1 + \lambda q^2, \nonumber \\
\lambda & = & (0.65 \pm 0.005) \; {\rm fm}^2.
\end{eqnarray}

\noindent The form factor $f_-$ does not contribute significantly to the
decay because its effect is proportional to $m_e$. The decay rate is

\begin{eqnarray}
BR(K_L \rightarrow \pi^0 e^+ e^- )_{\rm dir} = 4.16 ({\rm Im} \lambda_t )^2
(y^2_{7A} + y^2_{7V}), \nonumber \\
{\rm Im} \lambda_t = {\rm Im} V^{\phantom{*}}_{td}
V^*_{ts} = \vert V^{\phantom{*}}_{ub} \vert \vert V^{\phantom{*}}_{cb} \vert
\sin \delta = A^2 \lambda^5 \eta,
\end{eqnarray}

\noindent where $V^{\phantom{*}}_{td} = \vert V^{\phantom{*}}_{ub}
\vert \sin \delta$, and $A, \lambda,
\eta$ referred to the Wolfenstein parameterization of the KM matrix
\cite{W}. This results in

\begin{equation}
BR (K_L \rightarrow \pi^0 e^+ e^-)_{\rm dir} = \left\{ \begin{array}{l}
2.1 \times 10^{-12} \\
2.75 \times 10^{-12} \\
3.6 \times 10^{-12} \end{array} \right\} \times \left( {{\rm Im}
\lambda_t \over
10^{-4}} \right)^2 \; {\rm for} \; m_t = \left\{ \begin{array}{l}
150 \; {\rm GeV} \\
175 \; {\rm GeV} \\
200 \; {\rm GeV} \end{array} \right. .
\end{equation}

\noindent The dependence on the top quark mass will of course be removed
by a convincing precise measurement of $m_t$.

The KM parameter $V_{cb}$ has the most favored values (including the
recent CLEO data) \cite{S}

\begin{equation}
V^{\phantom{*}}_{cb} = \left\{ \begin{array}{ll}
0.036 \pm 0.002 \pm 0.002 \quad & {\rm HQET}, \\
0.036 \pm 0.002 \pm 0.003 & B \rightarrow D^{\ast} \ell \nu \; {\rm
models}, \\
0.039 \pm 0.001 \pm 0.004 & B \rightarrow X \ell \nu, \\
0.036 \pm 0.003 & ({\rm averaged}). \end{array} \right.
\end{equation}

\noindent The element $V^{\phantom{*}}_{ub}$ is measured by the
inclusive decay $B
\rightarrow X_u e \nu$ in the electron endpoint region. The two inclusive
calculations available yield

\begin{equation}
{V^{\phantom{*}}_{ub} \over V^{\phantom{*}}_{cb}} =
\left\{\begin{array}{ll}
0.082 \pm 0.006 \quad & \left[ {\rm ACCMM} \right] \quad\cite{ACCMM}, \\
0.074 \pm 0.007 & \left[ {\rm RDB} \right] \quad\cite{RDB}.
\end{array} \right.
\end{equation}

\noindent Models that calculate a set of exclusive decays ($B \rightarrow
Me \nu$) can only be used to provide an upper bound on
$V^{\phantom{*}}_{ub}$ since
there are many final states (such as $B \rightarrow \pi \pi e \nu$ with $\pi
\pi$
nonresonant) that are not calculated. These limits are

\begin{equation}
{V^{\phantom{*}}_{ub} \over V^{\phantom{*}}_{cb}} \leq
\left\{\begin{array}{ll}
0.12 \quad & [{\rm ISGW}] \quad\cite{ISGW}, \\
0.087 \quad & [{\rm BSW}] \quad\cite{WSB}, \\
0.067 \quad & [{\rm KS}] \quad\cite{KS}. \end{array} \right.
\end{equation}

\noindent We will use the former measurements to estimate

\begin{equation}
\left\vert {V^{\phantom{*}}_{ub} \over V^{\phantom{*}}_{cb}} \right\vert = 0.078 \pm 0.007 \pm 0.010,
\end{equation}

\noindent with the first uncertainty experimental and the second theoretical.
In the Wolfenstein parameterization of the KM matrix, the values of
$V^{\phantom{*}}_{cb}$ and $V^{\phantom{*}}_{ub}$ imply

\begin{eqnarray}
A & = & 0.74 \pm 0.06, \nonumber \\
\sqrt{\rho^2 + \eta^2} & = & 0.355 \pm 0.056.
\end{eqnarray}

\noindent Without any further analysis, these measurements imply an upper
bound on ${\rm Im} \lambda_t$:

\begin{eqnarray}
{\rm Im} \lambda_t & = & \vert V^{\phantom{*}}_{ub}
\vert \vert V^{\phantom{*}}_{cb} \vert \sin \delta
\nonumber \\
& = & (1.0 \pm 0.3) \times 10^{-4} \sin \delta.
\end{eqnarray}

\noindent A lower bound on this parameter can be found by consideration of
the analysis of $\epsilon$. In the Wolfenstein parameterizations one has the
approximate form

\begin{equation}
\vert \epsilon \vert = (3.4 \times 10^{-3}) A^2 \eta B_K \left[ 1 + 1.3 A^2
(1 - \rho) \left( {m_t \over m_W} \right)^{1.6} \right],
\end{equation}

\noindent where $B_K$ parameterizes the hadronic matrix element and is
estimated to be in the range $0.33 \leq B_K \leq 1$. Using $(1 - \rho) <
1.4$ and $B_K < 1$ one finds

\begin{equation}
A^2 \eta \geq \left\{\begin{array}{ll}
0.104 \quad & m_t = 200 \; {\rm GeV}, \\
0.13 & m_t = 175 \; {\rm GeV}, \\
0.16 & m_t = 150 \; {\rm GeV}, \end{array} \right.
\end{equation}

\noindent so that

\begin{equation}
{\rm Im} \lambda_t = A^2 \lambda^5 \eta \geq \left\{\begin{array}{l}
5.3 \times 10^{-5} \quad m_t = 200 \; {\rm GeV}, \\
6.8 \times 10^{-5} \quad m_t = 175 \; {\rm GeV}, \\
8.2 \times 10^{-5} \quad m_t = 150 \; {\rm GeV}. \end{array} \right.
\end{equation}

\noindent This brackets the range

\begin{equation}
0.53 \times 10^{-4} \leq {\rm Im} \lambda_t \leq 1.3 \times 10^{-4}.
\end{equation}

\noindent Note that ${\rm Im} \lambda_t$ is positive. These
constraints yield a
decay rate from direct CP violation of magnitude

\begin{eqnarray}
\left. \begin{array}{l} 1.01 \\ 1.25 \\ 1.4 \end{array} \right\}
\times 10^{-12} \leq BR (K_L \rightarrow \pi^0 e^+ e^-)_{\rm dir} \leq
\left\{ \begin{array}{l} 6.1 \\ 4.6 \\ 3.5 \end{array} \right\}
\times 10^{-12} \; {\rm for} \; m_t =
\left\{ \begin{array}{l} 200 \\ 175 \\ 150 \end{array} \right. .
\end{eqnarray}

\noindent Alternately, the ``best" values

\begin{eqnarray}
{\rm Im} \lambda_t & = & 1.0 \times 10^{-4}, \nonumber \\
m_t & = & 175 \; {\rm GeV},\end{eqnarray}

\noindent which we will take as our standard reference values, lead to a
rate

\begin{equation}
BR (K_L \rightarrow \pi^0 e^+ e^-)_{\rm dir} = 2.32 \times 10^{-12}.
\end{equation}

\noindent A more detailed analysis including a correlation between $\rho$
and $\eta$ inherent in Eq. (15), as well as the use of $B^0_d
\overline{B}^0_d$ mixing (which constrains $A \sqrt{(\rho - 1)^2 + \eta^2}$ as
a function of $f_B$) narrows the range only slightly because hadronic
uncertainties dominate.

\section{Revising the EPR analysis}

In this section, we review the formalism for analyzing mass matrix CP
violation, first set forth by Ecker, Pich, de Rafael (EPR)
\cite{EPR1,EPR2}. This amounts to
the prediction of the decay rate for $K_S \rightarrow \pi^0 e^+ e^-$, since
the mass matrix effect is defined by

\begin{eqnarray}
A (K_L \rightarrow \pi^0 e^+ e^-) \vert_{\rm MM} & \equiv & \epsilon A
(K_S \rightarrow \pi^0 e^+ e^-), \nonumber \\
\epsilon & = & (2.258 \times 10^{-3}) e^{i \pi / 4}.
\end{eqnarray}

\noindent We then redo the results taking into account recent work on the
nonleptonic kaon decays to one loop order. While this produces a
significant numerical change, it is more important as a prelude to our
subsequent analysis of uncertainties in the analysis.

The prediction of $K_S \rightarrow \pi^0 e^+ e^-$ comes from a
comparison with $K^+ \rightarrow \pi^+ e^+ e^-$, which contains many of
the same ingredients. The reactions are displayed schematically in Figs. 2,
3. In these diagrams the round circles represent the electromagnetic
coupling while the square boxes indicate the action of the weak interaction.
We know the electromagnetic interactions of pions and kaons from direct
measurement. The weak $K \rightarrow \pi$ transition of Fig. 2a,b is known
within some theoretical uncertainty from the use of chiral symmetry to relate
it to $K \rightarrow 2 \pi$ and $K \rightarrow 3 \pi$. However, the weak $K
\pi \gamma$ vertex is not known a priori and needs to be extracted from the
analysis of $K^+ \rightarrow \pi^+ e^+ e^-$.

\begin{figure}[t]
\centering
\leavevmode
\centerline{
\epsfbox{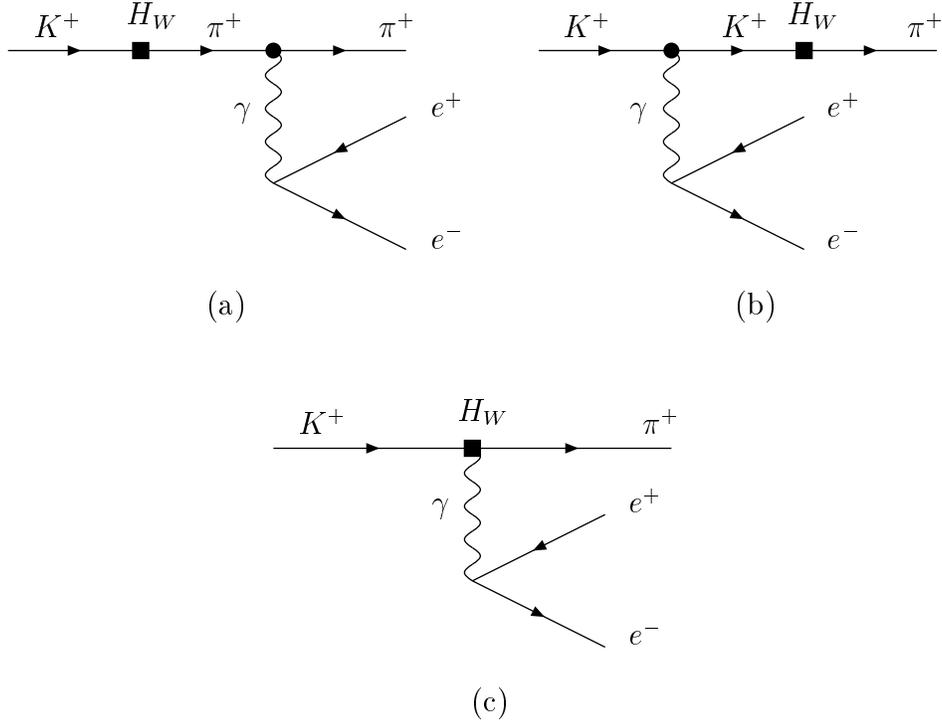}}
\caption{Diagrams contributing to the reaction $K^+ \ri
\pi^+ e^+ e^-$. Round circles represent the electromagnetic coupling
while the square boxes indicate the action of the weak interaction.}
\end{figure}

\begin{figure}[t]
\centering
\leavevmode
\centerline{
\epsfbox{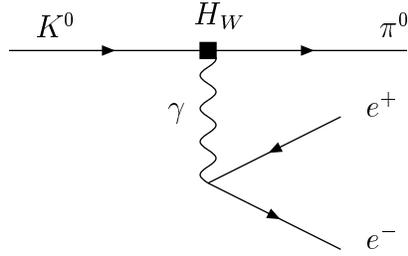}}
\caption{Same as in Fig. 2 for $K_L \ri \pi^0 e^+ e^-$.}
\end{figure}

The nonleptonic weak interactions are described by effective chiral
Lagrangians, organized in an expansion in powers of the energy, or
equivalently in numbers of derivatives and masses. At lowest order, called
order $E^2$, the physical transitions are described by a unique Lagrangian

\begin{eqnarray}
{\cal L} & = & G_8 Tr (\lambda_6 D_{\mu} U D^{\mu} U^{\dagger}) ,
\nonumber \\
U & \equiv & \exp \left(i {\lambda^A \cdot \phi^A \over
F_{\pi}}\right), \quad A = 1 \ldots 8,
\end{eqnarray}

\noindent where $\phi^A$ are the octet of pseudoscalar mesons ($\pi, K,
\eta$). At next order, order $E^4$, the number of possible forms of
Lagrangians is quite large, and has been categorized by Kambor, Missimer
and Wyler \cite{KMW1}. Not all of these contribute to $K \rightarrow 2 \pi$ and $K
\rightarrow \pi$, but certain linear combinations do influence these
amplitudes. The formalism of chiral perturbation theory dictate that when
an analysis is carried out to order $E^2$, that one use Eq. (23) at
tree level, in which case one obtains from the $K \rightarrow \pi$ decay rate

\begin{eqnarray}
G_8 & = & {G_F \over \sqrt{2}} \vert V^{\phantom{*}}_{ud} V^*_{us} \vert
g^{\phantom{\dagger}}_8, \nonumber \\
g^{\rm tree}_8 & = & 5.1.
\end{eqnarray}

Note that we neglect the CP violation in the nonleptonic
amplitude, contained in $G_8$, because this is bounded to be tiny by the
smallness of $\epsilon^{\prime}/\epsilon$.

\noindent In contrast, when evaluated at order $E^4$, one must include one
loop diagrams in addition to the possible order $E^4$ Lagrangian. The
loop diagrams involving $\pi \pi$ rescattering in the $I = 0$ channel, $K
\rightarrow (\pi \pi)_{I = 0} \longrightarrow (\pi \pi)_{I = 0}$ as pictured in
Fig. 4, are quite large and are the major part of the order $E^4$ analysis.
While there is some ambiguity in the extraction of
$g^{\phantom{\dagger}}_8$ (see below), the
enhancement from $\pi \pi$ rescattering leads to a smaller value of
$g^{\phantom{\dagger}}_8$, with a good estimate being \cite{KMW2}

\begin{equation}
g^{\rm loop}_8 = 4.3.
\end{equation}

\begin{figure}[!htbp]
\centering
\leavevmode
\centerline{
\epsfbox{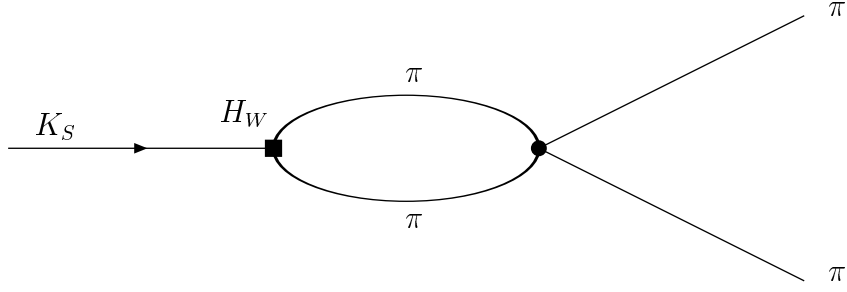}}
\caption{One-loop diagrams involving $ \pi \pi$ rescattering in
the $I$ = 0 channel for $K_S \ri ( \pi \pi )_{I = 0} \longrightarrow
( \pi \pi )_{I = 0}$. The notation for vertices is as in Fig. 2.}
\end{figure}

\noindent The $K \rightarrow \pi$ amplitude used in Figs. 2, 3 does not have
the enhancement from $\pi \pi$ rescattering, and is given in terms of
$g^{\phantom{\dagger}}_8$ by

\begin{eqnarray}
A(K^+ \rightarrow \pi^+) & = & 2 G_8 k^2, \nonumber \\
A(K^0 \rightarrow \pi^0) & = & - \sqrt{2} G_8 k^2.
\end{eqnarray}

\noindent We will explore further uncertainties in the $K \rightarrow \pi$
vertex in the next section.

To complete the diagrams of Figs. 2, 3 requires the electromagnetic
vertices of kaons and pions. In chiral perturbation theory to order $E^4$
these are given by

\begin{eqnarray}
\langle \pi^+ \vert J^{\mu} \vert \pi^+ \rangle & = & \left\{ 1 +q^2
\left[ {2 L_9 (\mu) \over F^2_{\pi}} - {1 \over 96 \pi^2 F^2_{\pi}} \left(\log
{m^2_{\pi} \over \mu^2} + {1 \over 2} \log {m^2_K \over \mu^2} + {3
\over 2} \right) \right] \right\} (p + p^{\prime})^{\mu}, \nonumber \\
\langle K^+ \vert J^{\mu} \vert K^+ \rangle & = & \left\{ 1 + q^2 \left[ {2
L_9 (\mu) \over F^2_{\pi}} - {1 \over 96 \pi^2 F^2_{\pi}} \left(\log
{m^2_K \over \mu^2} + {1 \over 2} \log {m^2_{\pi} \over \mu^2} + {3
\over 2} \right) \right] \right\} (p + p^{\prime})^{\mu}. \nonumber \\
\end{eqnarray}

\noindent The first of these is known more fully from experiment, and has
the form

\begin{eqnarray}
\langle \pi^+ \vert J^{\mu} \vert \pi^+ \rangle & = & {(p +
p^{\prime})^{\mu} \over [ 1 - q^2 / m^2 ]}, \nonumber \\
m & = & \; 730 \; {\rm MeV}.
\end{eqnarray}

\noindent Taylor expanding the latter form one determines $L_9
(\mu = m_{\eta}) = (7.4 \pm 0.7) \times 10^{-3}$. The experimental
charge radii

\begin{eqnarray}
\langle r^2 \rangle_{\pi^+} & = & (0.44 \pm 0.02) \; {\rm fm}^2, \nonumber \\
\langle r^2 \rangle_{K^+} & = & (0.34 \pm 0.05) \; {\rm fm}^2, \nonumber \\
\end{eqnarray}

\noindent are compatible with this value.
The final ingredient required for Figs. 2c, 3 is the weak photonic
coupling. This includes both short distance and long distance physics, as
illustrated in Fig. 5. While the short distance components have a reliable
hadronic matrix element (it is due to the real parts of the coefficients
discussed in the previous section), the QCD coefficient depends strongly on
the low energy cutoff $\mu$. In the full matrix element, this dependence is
canceled by a corresponding dependence on $\mu$ of the long distance
physics. However, since the long distance physics is not calculable, we
must attempt to determine this coupling phenomenologically. The
innovation
of EPR was to elucidate the possible forms that this coupling could take.
They found that there were two possible chiral Lagrangians which could
contribute to this process:

\begin{eqnarray}
{\cal L}_W & = & {i e G_8 \over 2} F^{\mu \nu} \left[ w_1 Tr \left( Q
\lambda_{6-i7} {\cal L}_{\mu} {\cal L}_{\nu} \right) + w_2 Tr \left( Q
{\cal L}_{\mu} \lambda_{6-i7} {\cal L}_{\nu} \right) \right], \nonumber \\
{\cal L}_{\mu} & \equiv & - \left( \partial_{\mu} U - i e [A_{\mu}, U]
\right) U^{\dagger} = -(D_{\mu} U) U^{\dagger}.
\end{eqnarray}

\noindent In the presence of the short distance electroweak penguin effect
due to Z exchange with an axial electron coupling, we need a third effective
Lagrangian, not present in EPR,

\begin{equation}
{\cal L}^{\prime}_W = {i 2 \pi \alpha \over 3} G_8 w_5 \bar{e}
\gamma^{\mu} \gamma_5 e Tr \left( \lambda_{6-i7}{\cal L}_{\mu} \right).
\end{equation}

\noindent The correspondence with the notation of the last section is

\begin{equation}
{\rm Im} \; w_5 = {3 \over 4 \pi} {1 \over {\vert V^{\phantom{*}}_{ud}
V^*_{us} \vert g^{\phantom{\dagger}}_8}} {y^{\phantom{\dagger}}_{7A}
\over \alpha} {\rm Im} \lambda_t
\end{equation}

\begin{figure}[!htbp]
\centering
\leavevmode
\centerline{
\epsfbox{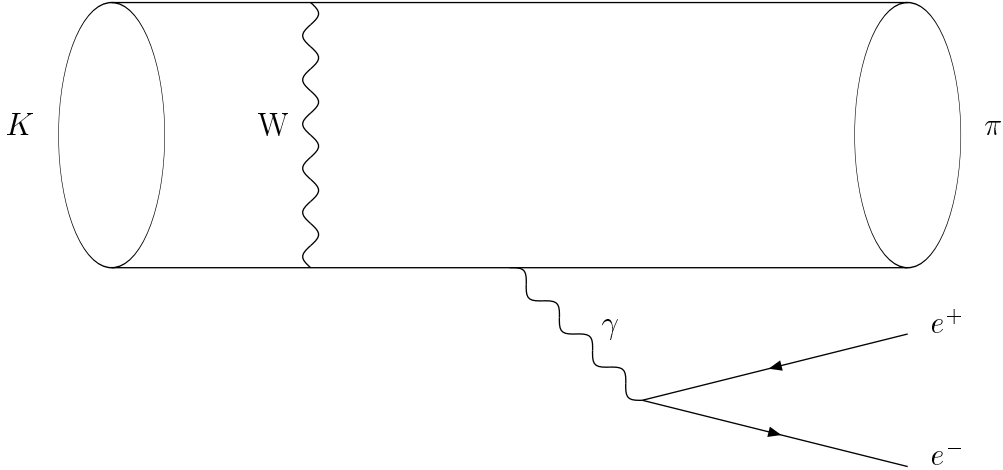}}
\caption{Diagrams contributing to the short distance weak photonic
coupling for $K \ri \pi e^+ e^-$.}
\end{figure}

\noindent Note that the labeling of $w_5$ takes into account the Lagrangians
labeled by $w_3$, $w_4$ defined by EPR \cite{EPR1,EPR2,EPR3} which
contributed to other radiative $K$
decays. In fact this form of ${\cal L}^{\prime}_W$ is closely related to $w_1$
term since by using identities of the U matrix it can be shown that

\begin{eqnarray}
F^{\mu \nu} Tr (Q \lambda_{6-i7} {\cal L}_{\mu} {\cal L}_{\nu} ) & = &
-F^{\mu \nu} \partial_{\nu} Tr (Q \lambda_{6-i7} {\cal L}_{\nu} )
\nonumber \\
& = & (\partial_{\mu} F^{\mu \nu} ) Tr(Q \lambda_{6-i7} {\cal L}_{\nu}
) \nonumber \\
& = & (-e) \bar{e} \gamma_{\mu} e Tr(Q \lambda_{6-i7} {\cal L}_{\nu} )
\nonumber \\
& = & \left( {e \over 3} \right) \bar{e} \gamma_{\mu} e Tr (\lambda_{6-
i7} {\cal L}_{\nu}),
\end{eqnarray}

\noindent where in the second line we have integrated by parts thus
subsequently used the equation of motion so that

\begin{equation}
{i e G_8 \over 2} F_{\mu \nu} w_1 Tr (Q \lambda_{6-i7} {\cal L}_{\mu}
{\cal L}_{\nu} ) = {i \pi \alpha G_8 \over 3} w_1 \bar{e} \gamma_{\mu} e
Tr ( \lambda_{6-i7} {\cal L}_{\nu} ).
\end{equation}

\noindent This allows us to identify the short distance CP-violating part of
$w_1$:

\begin{equation}
{\rm Im} \; w_1 = {3 \over 4 \pi} {1 \over {\vert V^{\phantom{*}}_{ud}
V^*_{us} \vert g^{\phantom{\dagger}}_8}} {y^{\phantom{\dagger}}_{7V} \over
\alpha} {\rm Im} \lambda_t.
\end{equation}

The real parts of $w_1, w_2$ contain long distance contributions and
hence are not
predictable by present techniques. They need to be extracted from
experimental measurements. The other process available for this procedure
is $K^+ \rightarrow \pi^+ e^+ e^-$ (unless $K_S \rightarrow \pi^0 e^+ e^-
$ is measured in the future). Unfortunately one cannot fix both $w_1,
w_2$ in this way, so that one must add a theoretical assumption in order to
proceed. The relevant amplitudes are

\begin{eqnarray}
{\cal M}(K^+ \rightarrow \pi^+ e^+ e^-) & = &\phantom{-}{G_8\alpha \over 4 \pi}
d_+ (p + p^{\prime})^{\mu} \bar{u} \gamma_{\mu} v, \nonumber \\
{\cal M}(K_S \rightarrow \pi^0 e^+ e^-) & = & - {G_8 \alpha \over 4 \pi}
d_S (p + p^{\prime})^{\mu} \bar{u} \gamma_{\mu} v, \nonumber \\
{\cal M}(K_L \rightarrow \pi^0 e^+ e^-) & = & - {G_8 \alpha \over 4 \pi} (p +
p^{\prime})^{\mu} \bar{u} \left[ d_V \gamma_{\mu} + d_A \gamma_{\mu}
\gamma_5 \right] v,
\end{eqnarray}

\noindent with

\begin{eqnarray}
d_+ & \equiv & w_+ + \phi_K (q^2) + \phi_{\pi} (q^2), \nonumber \\
d_S & \equiv & {\rm Re} \; w^{\phantom{2}}_S + 2 \phi_K (q^2), \nonumber \\
w_+ & = & - {16 \pi^2 \over 3} \left( w^r_1 + 2 w^r_2 - 12 L^r_9
\right) - {1 \over 6} \log {m^2_K m^2_{\pi} \over m^4_{\eta}}, \nonumber\\
w^{\phantom{2}}_S & = & w_+ + 16 \pi^2 \left( w^r_2 - 4 L^r_9 \right)
+ {1 \over 6} \log {m^2_{\pi} \over m^2_K}, \nonumber \\
\phi_i (q^2) & = & {m^2_i \over q^2} \int^1_0 dx \left[ 1 - {q^2 \over
m^2_i} x (1 - x) \right] \log \left[ 1 - {q^2 \over m^2_i} x (1 - x) \right],
\end{eqnarray}

\noindent and for the CP-violating $K_L$ decay

\begin{eqnarray}
d_V & = & \epsilon d_S - {16 \pi^2 \over 3} i {\rm Im} \; w_1, \nonumber \\
d_A & = & -{16 \pi^2 \over 3} i {\rm Im} \; w_5.
\end{eqnarray}

\noindent The goal of the search for direct CP violation is to separate the
${\rm Im} \; w_{1,5}$ terms from the mass matrix effect $\epsilon
d_S$. Note that
in these expressions we have neglected the possible direct CP violation in
the $K \rightarrow \pi$ transition (which is bounded to be very small by the
measurement of $\epsilon^{\prime} / \epsilon$) and the contribution of
${\rm Re} \; w_5$
to CP-conserving decays (since ${\rm Re} \; w_5 \approx {\rm Im} \;
w_5 << {\rm Re} \; w_1$).

The EPR analysis of $K^+ \rightarrow \pi^+ e^+ e^-$ uses the tree
level value of $g^{\phantom{2}}_8$, $g^{\rm tree}_8 = 5.1$. The decay rate is
consistent with two values of ${\rm Re} \; w_+$, and a subsequent
analysis of the
decay spectrum favored the lower value for ${\rm Re} \; w_1$, i.e.
${\rm Re} \; w_+$ = 1.16
$\pm$ 0.08 \cite{EPR1}. However, given that one is working to one loop order, it is more
consistent to use the one loop value for $g^{\phantom{2}}_8$,
$g^{\rm loop}_8$ = 4.3. Because of the presence of the
$L_9$ term, this is not just a rescaling of the value of $w_+$. An
additional change that we make is to use the known full electromagnetic
vertex in the pole diagrams, rather than just the first term in the
expansion of the form factor. Note that because of the factor of $k
\cdot p$ in the weak matrix element, the only significant form factor
is that of the pion in Fig. 2. This implies the replacement

\begin{equation}
{2 L^r_9 \over F^2_{\pi}} \rightarrow {2L_9^r \over F^2_{\pi} \left( 1 - {q^2
/ m^2_{\rho}} \right)}
\end{equation}

\noindent in the formula for $w_+$. The associated logarithm with $\mu
\approx m_{\eta}$ are much smaller than the $L^r_9$ dependence and are
not influenced much by this replacement. As a technical note, we
comment that some potential modifications using a phenomenological
pion form factor could lead to a lack of gauge invariance. By
modifying the coefficient of a gauge invariant effective Lagrangian,
we preserve the gauge invariant nature of the amplitude. With these
changes, we find

\begin{equation}
{\rm Re} \; w_+ = 1.01 \pm 0.10.
\end{equation}

\noindent [Without the second change, we would have had ${\rm Re} \; w_+ = 1.33
\pm 0.065$]. This is illustrated in Fig. 6.

\begin{figure}[!htbp]
\centering
\leavevmode
\epsfysize=200pt
{\centerline{\epsfbox{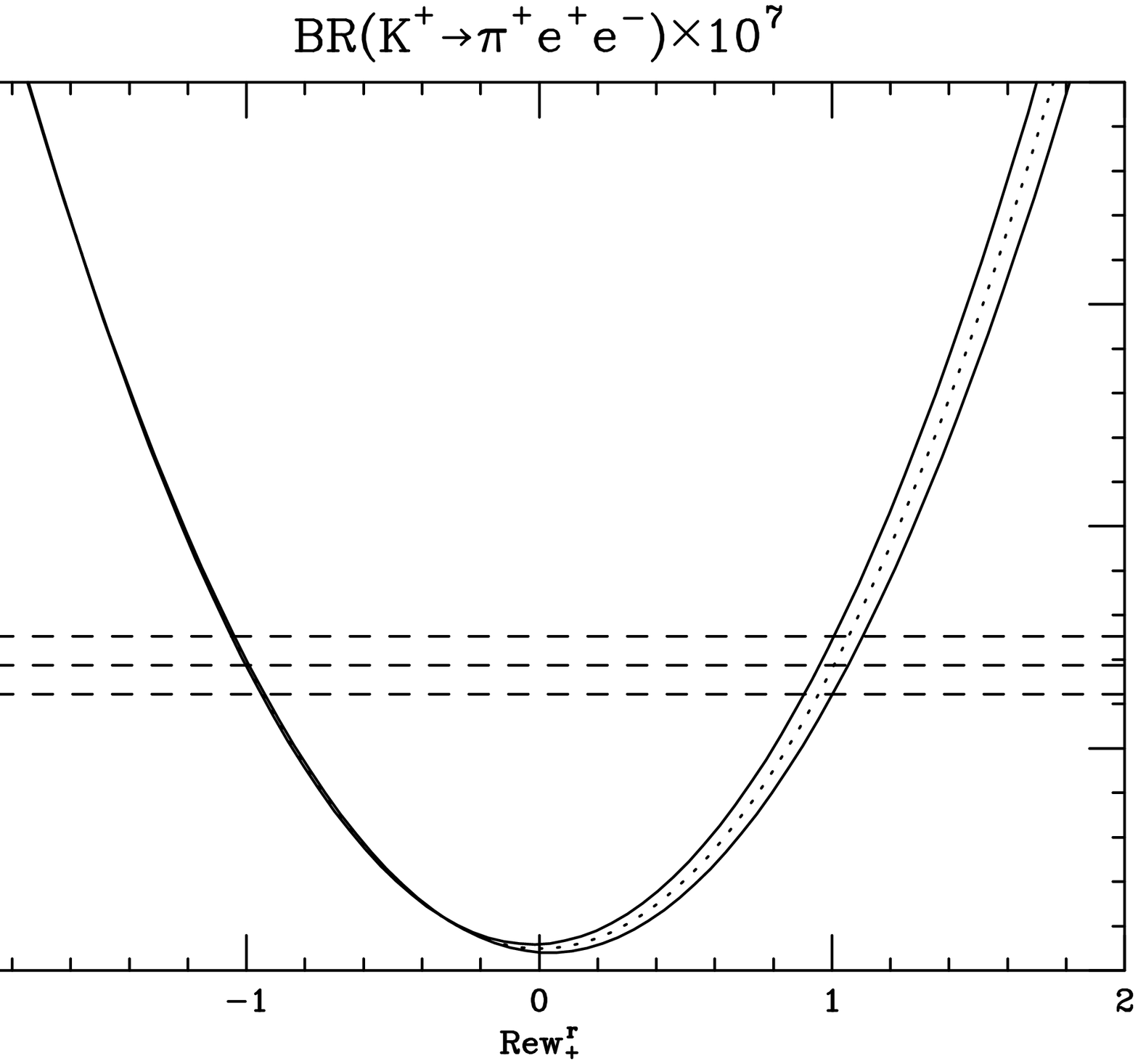}}}
\caption{The branching ratio $BR(K^+ \ri \pi^+ e^+ e^-)$ is
plotted against ${\rm Re} \; w_+$. The solid curves are obtained using
the extreme values of the error intervals of
$\vert V_{ud} \vert$ = 0.9753 $\pm$ 0.0006,
$\vert V_{us} \vert$ = 0.221 $\pm$ 0.003
and $L_9(m_{\eta})$ = (7.4 $\pm$ 0.7)$\times$ $10^{-3}$, while the dashed
curve corresponds to the central values. The experimental value of the
branching ratio $\pm$ its experimental errors are indicated by dashed
horizontal lines.}
\end{figure}

One cannot simply transfer this information to $K_S$ or $K_L$ decays,
because a different linear combination enters

\begin{equation}
w^{\phantom{2}}_S = w_+ + 16 \pi^2 ( w_2 - 4 L_9) + {1 \over 6}
\log {m^2_{\pi} \over m^2_K}.
\end{equation}

\noindent However EPR deal with this problem by making the assumption
that $w_2 = 4L_9$ resulting in ${\rm Re} \; w^{\phantom{2}}_S = 0.73
\pm 0.08$ \cite{EPR1}.
They note that this
is an assumption which is not part of chiral perturbation theory, but do not
explore the consequences if it is not correct. We will discuss this
in the next
section, finding a very strong sensitivity. At this stage we note that if one
makes the assumption of $w_2 = 4 L_9$, one obtains ${\rm Re} \; w^{\phantom{2}}_S
= 0.58 \pm 0.10$ for our value of ${\rm Re} \; w_+$.

At this value of $w^{\phantom{2}}_S$, the direct and mass matrix
contributions are comparable:

\begin{eqnarray}
d_V & = & \epsilon d_S - {16 \pi^2 \over 3} i {\rm Im} \; w_1 \nonumber \\
& \approx & e^{i {\pi / 4}} \left( 0.57 \times 10^{-3} \right)
- i 1.0 \times 10^{-3}, \nonumber \\
d_A & = & i 1.0 \times 10^{-3},
\end{eqnarray}

\noindent when evaluated with $m_t = 175$ GeV, ${\rm Im} \lambda_t =
10^{-4}$. This leads to a branching ratio

\begin{equation}
BR ( K_L \rightarrow \pi^0 e^+ e^-)_{\rm MM} = 0.37 \times 10^{-12}
\end{equation}

\noindent if there is no direct CP violation $({\rm Im} \lambda_t = 0)$ [EPR
found $BR_{\rm MM} = 1.5 \times 10^{-12}$ in this case \cite{EPR2}], vs.

\begin{equation}
BR ( K_L \rightarrow \pi^0 e^+ e^-)_{\rm CP} = 1.78 \times 10^{-12}
\end{equation}

\noindent for the full set of parameters given above in Eq. (37). The addition
of mass matrix CP violation in this analysis led to a small decrease in the
rate compared to the purely direct CP violation of the previous section,
Eq. (21), because of the cancellation in the imaginary part of $d_V$. However
this may change if $w_2 \neq 4L_9$.

One of the results that we will see in the next section is that the
mass matrix contribution to the branching ratio is near a minimum
value when $w_2$ is close to $4L_9$. For other values of $w_2$, the
rate can easily be an order of magnitude larger. Although our value
for the branching ratio is a factor of four below that of EPR, both
estimates are similar in saying that the mass matrix contribution will
be small as long as $w_2$ = $4L_9$.

Using the tree level value for $g^{\phantom{\dagger}}_8$, $g^{\rm
tree}_8$ = 5.1, the mass matrix contribution to the branching
ratio is

\begin{equation}
BR ( K_L \rightarrow \pi^0 e^+ e^-)_{\rm MM} = 0.55 \times 10^{-12}.
\end{equation}

\section{Uncertainties in mass matrix CP violation}

Our goal in this section is to assess how well we understand the prediction
for $K_S \rightarrow \pi^0 e^+ e^-$. The most important effect will be
discussed in section (c) below, but we proceed systematically to discuss
even contributions which have less uncertainty.

\begin{description}
\item[a)] {\bf Purely electromagnetic vertices.}

The electromagnetic vertices enter in diagrams 2, 3a,b. The
uncertainty here is in the choice of whether to use the chiral
expansion of the form factor truncated at order $q^2$, Eq. (27), or
the full $q^2$ dependence of the monopole form factor, Eqs. (28),
(39). The first choice is natural when one is working to a given order
in the chiral energy expansion, but the latter choice clearly includes
more of the physics which is known about the electromagnetic vertex.
Note that it is essentially only the pion form factor which is
relevant, because the diagram involving the kaon form factor, Fig.
2b, is suppressed by a factor of $m^2_{\pi}/m^2_K$ with respect to
Fig. 2a because of the momentum dependence of the weak $K \ri \pi$
transition. The use of the full form factor produces a modest
variation in the value of $w_+$ (i.e. $Re\; w_+ = 1.01$ instead of
$Re\; w_+ = 1.33$). Because of cancellations in the $K^0$ amplitude
this provokes a more extreme variation on the decay rate. The lowest order
chiral vertex, Eq. (27), produces a decay rate

\begin{equation}
BR ( K_L \rightarrow \pi^0 e^+ e^-)_{\rm MM} = 1.96 \times 10^{-12}
\end{equation}

\noindent instead of the result of Eq. (43). [We note that if we had
also modified $w_2$ in the same way as $4L_9$ as in Eq. (39) we would
have a $BR ( K_L \rightarrow \pi^0 e^+ e^-)_{\rm MM} = 1.33 \times
10^{-12}$.] While we feel that it is good physics to use the full
electromagnetic form factor, one could also interpret these results as
an uncertainty in the analysis due to higher order terms in $q^2$,
with that uncertainty being of order $2 \times 10^{-12}$.

\item[b)] {\bf The weak $K \rightarrow \pi$ vertex.}

We have already given one indication of the sensitivity of the result
to the size of the $K \rightarrow \pi$ transition. Under otherwise
identical assumptions, $g^{\rm tree}_8 = 5.1$ yielded the rate in Eq.
(45), while $g^{\rm loop}_8 = 4.3$ produced the result of Eq. (43).
These modifications to $g^{\phantom{l}}_8$ also include corresponding
modifications to the $K\pi\gamma$ vertex required by gauge invariance.
This is automatically maintained, however, by the use of gauge
invariant effective Lagrangians. In order to appreciate that this
change in $g^{\phantom{l}}_8$ is not the only uncertainties in the $K
\pi$ amplitude, one needs to understand a bit more about the chiral
phenomenology of $K \rightarrow 2 \pi$ and $K \rightarrow 3 \pi$.

Chiral symmetry relates processes with different numbers of pions, such as
$K \rightarrow \pi$ vs. $K \rightarrow 3 \pi$. The predictions are
compactly
contained in the chiral Lagrangians, but can also be obtained using the soft
pion theorems, which was the methodology used in the 1960's. The only
advantage of the latter technique is that it relies only on chiral SU(2) while
modern chiral Lagrangian analyses have always involved chiral SU(3)
symmetry. [Presumably the latter could be reformulated in chiral SU(2),
but no one has yet done this.] The soft pion analysis indicates that one
obtains the same relation (up to terms of order $m^2_{\pi}$) between $K
\rightarrow 2 \pi$ and $K \rightarrow 3 \pi$ for any Lagrangian which
survives in {\em any} soft pion limit of $K \rightarrow 3 \pi$ (i.e., $p_i
\rightarrow 0$). The only Lagrangians which do not survive in any soft
pion limit involve four separate derivatives on the four fields of $K
\rightarrow 3 \pi$, e.g.

\begin{equation}
{\cal L}_{\rm quartic} = {g^{\phantom{2}}_8 \over \Lambda^2_1}
Tr \left( \lambda_6 D_{\mu} U D_{\nu} U^{\dagger} D^{\mu} U
D^{\nu} U^{\dagger} \right).
\end{equation}

\noindent This Lagrangian yields a matrix element proportional to $(k \cdot
p_1) (p_2 \cdot p_3)$ which clearly vanishes as any $p_i \rightarrow 0$. In
contrast most of the order $E^4$ Lagrangians do not vanish in all soft pion
limits. An example is

\begin{equation}
{\cal L}^{\prime} = {g^{\phantom{2}}_8 \over \Lambda^2_2}
Tr \left( U^{\dagger} \lambda_6 D_{\mu} D_{\nu} U
D^{\mu} U^{\dagger} D^{\nu} U \right),
\end{equation}

\noindent which then yields the same relations of $K \rightarrow 3 \pi$ and
$K \rightarrow 2 \pi$ as does the lowest order result ${\cal L}$ given
in Eq. (23). [This phenomenon is explained in more detail in Ref. \cite{DGH}.]
Since the only inputs to
the chiral phenomenology are the amplitude for $K \rightarrow 2 \pi$ and
$K \rightarrow 3 \pi$, it follows that one cannot distinguish a combination
${\cal L} + {\cal L}^{\prime}$ from a Lagrangian involving ${\cal L}$ only.

\noindent However, when we discuss the $K \rightarrow \pi$ vertex,
there {\em is} a distinction between these various Lagrangians.
For example ${\cal L}^{\prime}$
in Eq. (48) involves a minimum of three meson fields, and hence contributes to
$K \rightarrow 2 \pi$ and $K \rightarrow 3 \pi$ but not at all to $K
\rightarrow \pi$. In contrast, the lowest order Lagrangian Eq. (23) contributes
to all of $K \rightarrow \pi, K \rightarrow 2 \pi, K \rightarrow 3 \pi$. Since
known phenomenology cannot distinguish between linear combinations of
${\cal L} + {\cal L}^{\prime}$, this manifests itself in an
uncertainty in the
$K \rightarrow \pi$ vertex. Since the higher order Lagrangians are known
to
make a 25\% difference in relation between $K \rightarrow 2 \pi$ and $K
\rightarrow 3 \pi$, it would be unreasonable to take this uncertainty in $K
\rightarrow \pi$ to be any less than 25-30\%. Using
$g^{\phantom{\dagger}}_8 = 4.3 \cdot(1 \pm 30\%)$ yields a range

\begin{equation}
BR(K_L \rightarrow \pi^0 e^+ e^-)_{\rm MM} \sim {\cal O}(10^{-16}) \div
1.5 \times 10^{-12}.
\end{equation}

\noindent This again indicates that the analysis has significant
cancellations present and that modest variations in the analysis can
lead to uncertainties of order 1.5$\times 10^{-12}$.

\item[c)] {\bf The $K \pi \gamma$ vertex.}

The parameter $w_+$ was determined from the analysis of $K^+
\rightarrow \pi^+ e^+ e^-$. This arrangement has some intrinsic
uncertainty because it was performed to a given order in the chiral energy
expansion. It would be reasonable to take this uncertainty at 30\%.

However, this uncertainty is dwarfed by the errors introduced by the
assumption of $w_2 = 4 L_9$. There is nothing about chiral
symmetry which forces such a relation. For example, if the process of Fig.
7 were to contribute to the weak coupling, the relation $w_2 = 4 L_9$
would not occur except for a special value of the $K
\rightarrow a_1, f_1$ amplitude. This is highly unlikely, and
we can easily accept $w_2 - 4 L_9 \neq 0$. The crucial distinction
here
is between models and rigorous theory. Chiral perturbation theory is a
rigorous method which expresses true relationships in QCD to a given order
in the energy expansion. However an assumption such as $w_2 = 4
L_9$ may be true or false in a way that we cannot decide based on QCD. It
may occur within some models, yet we have no guidance as to whether it is
correct in nature. We cannot base something as important as the
observation of direct CP violation on something as flimsy as a model.

\begin{figure}[!htbp]
\centering
\leavevmode
\centerline{
\epsfbox{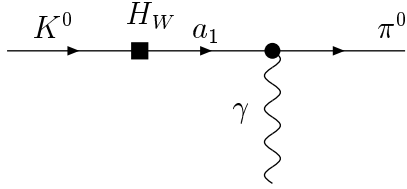}}
\vspace{0.3cm}
\caption{Diagram contributing to the weak photon coupling.
The notation for the vertices is as in Fig. 2.}
\end{figure}

Unfortunately the decay rate for $K_S \rightarrow \pi^0 e^+ e^-$ depends
very strongly on the value of $w_2$. This strong dependence was
observed by Littenberg and Valencia \cite{LV}. If we were to chose $w_2 = 0$,
the rate would be two orders of magnitude larger. In Fig. 8 we plot the
branching ratio for the mass matrix contribution to $K_L \rightarrow \pi^0
e^+ e^-$ vs. $w_2$. We see that reasonable values of $w_2$,
we get a wide range of values of the branching ratio. Conversely, a
measured value of $BR (K_L \rightarrow \pi^0 e^+ e^-)$ in the range of
$10^{-12} \rightarrow {\rm few} \times 10^{-11}$ could be interpreted in terms
of a reasonable value of $w_2$. This is then an enormous uncertainty
in the mass matrix contributions to $K_L \rightarrow \pi^0 e^+ e^-$.

This uncertainty could be removed if one measured the rate of $K_S
\rightarrow \pi^0 e^+ e^-$. The mass matrix contribution would then be
known, see Eq. (22). However this task is not easy experimentally.

\end{description}

\begin{figure}[!htbp]
\centering
\leavevmode
\epsfysize=200pt
{\centerline{\epsfbox{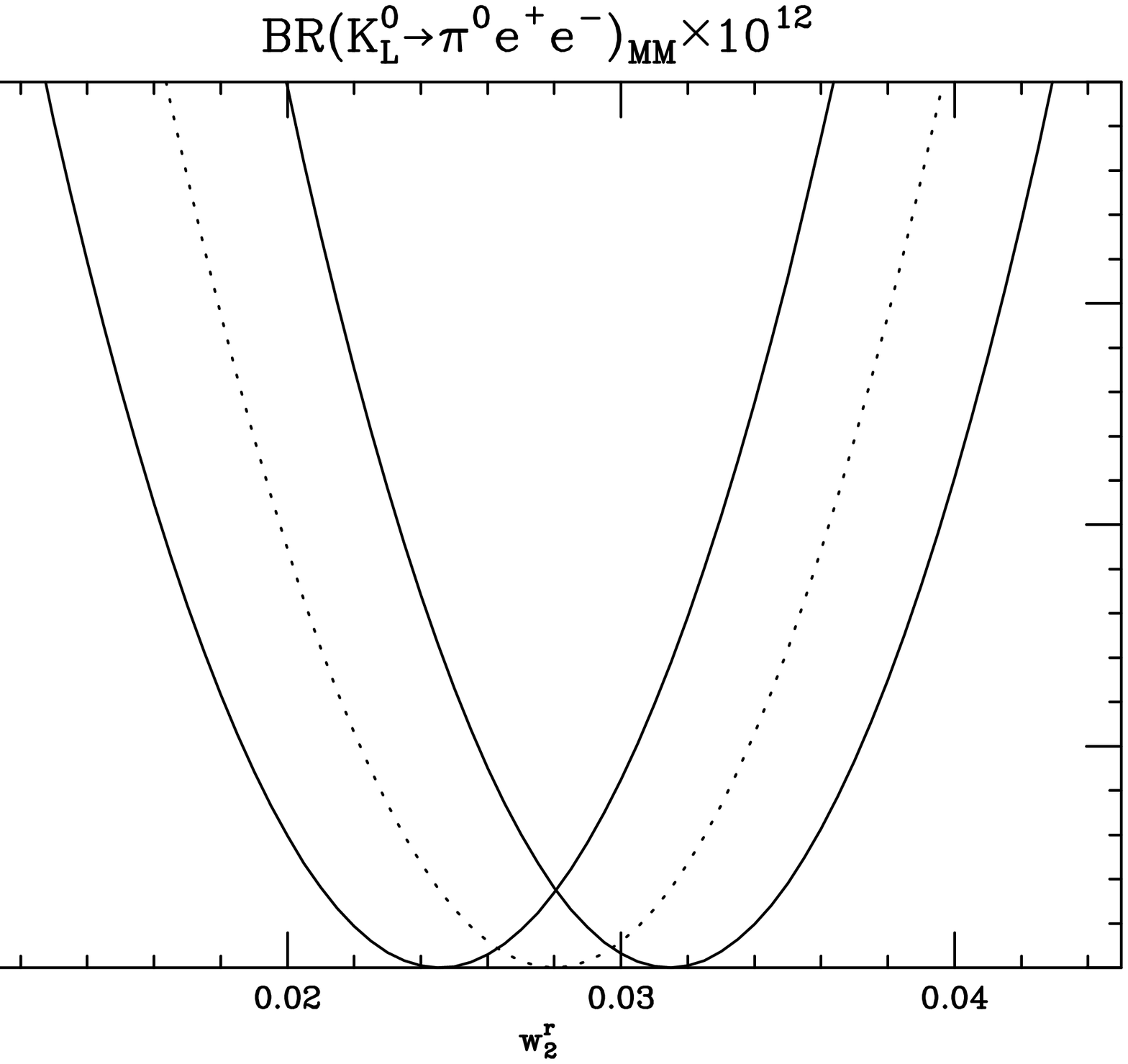}}}
\caption{The mass matrix contribution for the branching
ratio $BR(K_L \ri \pi^0 e^+ e^-)_{\rm MM}$ is plotted against $w_2$. The
convention for the solid and dashed curves is the same as for Fig. 6.}
\end{figure}

\section{The CP-conserving amplitude}

The $K_L \rightarrow \pi^0 e^+ e^-$ transition can also take place through
a CP-conserving two-photon intermediate state. If we ignore the electron
mass, the form of the amplitude will be

\begin{equation}
{\cal M} (K_L \rightarrow \pi^0 e^+e^-)_{\rm CPC} =
G_8 \alpha^2 K p \cdot (k_{e^+} - k_{e^-}) (p + p^{\prime})^{\mu}
\bar{u} \gamma_{\mu} v,
\end{equation}

\noindent where $K$ is a form factor and the extra
antisymmetry under $k_{e^+} \leftrightarrow k_{e^-}$ is a reflection of the
properties under a CP transformation. In order to calculate this we need to
understand the $K_L \rightarrow \pi^0 \gamma \gamma$ transition first.

We are fortunate that $K_L \rightarrow \pi^0 \gamma \gamma$ is accessible
to present experiments, and significant phenomenology has been performed
on this reaction. We will utilize the work of Cohen, Ecker and Pich
\cite{CEP} as
representative of present work, with the understanding that future
experimental and theoretical work will clarify the analysis considerably.
The $K_L \rightarrow \pi^0 \gamma \gamma$ amplitude has two form factors
$A$, $B$, defined via

\begin{eqnarray}
{\cal M} \left( K_L \rightarrow \pi^0 \gamma \gamma \right) =
{G_8 {\alpha} \over 4 \pi} \epsilon_{\mu} (q_1) \epsilon_{\nu} (q_2) \left[
A \left( q^{\mu}_2 q^{\nu}_1 - q_1 \cdot q_2 \cdot g^{\mu \nu} \right)
\phantom{{B \over m^2_K}} \right. \nonumber \\
\left. +2 {B \over m^2_K} \left( k \cdot q_1 q^{\mu}_2 p^{\nu} + p
\cdot q_2 q^{\nu}_1 p^{\mu} - q_1 \cdot q_2 p^{\mu} p^{\nu} - g^{\mu
\nu} p \cdot q_1 p \cdot q_2 \right) \right].
\end{eqnarray}

\noindent When the photons couple to $e^+ e^-$, as in Fig. 9, it is well
known that the $A$ amplitude contributes to $K_L \rightarrow \pi^0 e^+ e^-$
only proportional to $m_e$, which is a small effect which we will drop. It
is the $B$ amplitude which is important for the $e^+ e^-$ final state. The
authors use a representation which fits the known $K_L \rightarrow \pi^+
\pi^- \pi^0$ amplitude in a dispersive treatment of $K_L \rightarrow \pi^0
\gamma \gamma$ and find

\begin{eqnarray}
B(x) & = & c_2 \left\{ {1 \over x} F(x) + {4 \over 3} ( 5 - 2x) \left[ {1 \over
6} + R (x) \right] + {2 \over 3} \log {m^2_{\pi} \over m^2_{\rho}} \right\} +
\beta - 8 a_V, \nonumber \\
x & = & { \left( k_{e^-} + k_{e^+} \right)^2 \over 4 m^2_{\pi}} ,
\nonumber \\
\beta & = & -0.13, \nonumber \\
c_2 & = & 1.11, \nonumber \\
F(x) & = & 1 - {1 \over x} \left[ \sin^{-1} \left( \sqrt{x} \right) \right]^2,
\, \qquad x \leq 1, \nonumber \\
& = & 1 + {1 \over 4x} \left[ \log {1 - \sqrt{1 - 1/x} \over 1 + \sqrt{1 -
1/x}} + i \pi \right]^2, \, \qquad x \geq 1, \nonumber \\
R(x) & = & - {1 \over 6} + {1 \over 2x} \left[ 1 - \sqrt{1/x - 1} \sin^{-1}
\left( \sqrt{x} \right) \right], \, \qquad x \leq 1, \nonumber \\
& & - {1 \over 6} + {1 \over 2x} \left[ 1 + {\sqrt{1 - 1/x} \over 2}\left( \log {1 -
\sqrt{1 - 1/x} \over 1 + \sqrt{1 - 1/x}} + i \pi \right) \right], \,
\qquad x \geq 1.
\end{eqnarray}

\begin{figure}[!htbp]
\centering
\leavevmode
\centerline{
\epsfbox{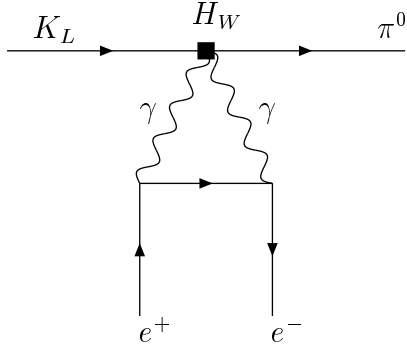}}
\caption{CP-nonviolating diagram involving two photons
coupling to $e^+ e^-$. The notation for the vertices is as in Fig. 2.}
\end{figure}

\noindent The most important ingredient above is $a_V$ which is an
unknown parameter representing the vector meson exchange contributions
to the $B$ amplitude. A fit to the decay rate (using $BR(K_L \rightarrow
\pi^0 \gamma \gamma) = 1.7 \times 10^{-6}$ \cite{NA}) and $\gamma
\gamma$ spectrum in $K_L
\rightarrow \pi^0 \gamma \gamma$ indicates a value around $a_V = -0.96$.
This parameter was very important in increasing the chiral prediction of the
decay rate to be in agreement with experiment. We have explored the
sensitivity of this parameter to changes in the analysis and have found that
25\% changes in the dispersive treatment lead to a factor of 2 change in
$a_V$, so that this value is still quite uncertain.

Several authors \cite{EPR3,CEP,PH} have calculated the contribution of
the on-shell two-photon
intermediate state to $K_L \rightarrow \pi^0 e^+ e^-$. Although this is
sometimes referred to as the absorption contribution, it is not the full
absorption part since there is a further cut due to on-shell pions. Besides
this, the full CP-conserving amplitude
also receives contribution form the dispersive part of the amplitude,
with off-shell photons (and pions). The calculation of this is complicated by the
sensitivity of the loop integral to high momentum, as the Feynman diagram
of Fig. 9 will diverge if we treat the $B$ amplitude as a constant in $q^2_1$
and $q^2_2$. However the remedy to this is well known; the couplings of
virtual photons to hadrons is governed by form factors which lead to
suppression of the couplings at high $q^2$. We will include an estimate of
these form factors and this will allow us to calculate the dispersive
component of the CP-conserving amplitude.

The two-photon loop integral in the limit $m_e \rightarrow 0$ is given by

\begin{eqnarray}
{\cal M} (K_L \rightarrow \pi^0 e^+ e^-)_{\rm CPC} = {G_8 \alpha \over 4 \pi
m^2_K} \int {d^4 \ell \over (2 \pi )^4} {B(k + k^{\prime}) F(\ell + k)
F( \ell - k^{\prime})
\over \ell^2 ( \ell + k) ^2 ( \ell - k^{\prime})^2} \times \nonumber \\
\left\{ \rlap/{p} \left[ \ell^2 + p \cdot (k - k^{\prime})
- p \cdot \ell \ell \cdot (k - k^{\prime})
+ 2p \cdot \ell k \cdot k^{\prime} - p \cdot k \ell \cdot k^{\prime} -
p \cdot k^{\prime} \ell \cdot k \right] + \right. \nonumber \\
\left. \rlap/{\ell} \left[ ( p \cdot \ell )^2 + p
\cdot \ell p \cdot (k - k^{\prime}) \right] \right\}.
\end{eqnarray}

\noindent Here $F(q^2)$ are the form factors for the virtual photon
couplings. The structure above is certainly an approximation, as in general
the virtual photon dependence need not be only an overall factor of
$F(q^2)$. However the above form would be sufficient to capture the
kinematic variation if only one photon is off-shell (given an appropriate
$F(q^2)$). Since we only need a minor form factor suppression to tame the
logarithmic divergences, we feel that this structure will be sufficient for our
estimate. We choose

\begin{equation}
F(q^2) = {-m^2_V \over (q^2 - m^2_V)},
\end{equation}

\noindent which is a good representation of almost any mesonic form factor,
with $m^{\phantom{\dagger}}_V \approx m_\rho$. Neglecting terms which
are suppressed by powers of $1/m^2_\rho$, we find the amplitude of Eq.
(50) above, with

\begin{equation}
K = {B(x) \over {16 \pi^2 m^2_K}} \left [
{2 \over 3} \log \left({m^2_{\rho}} \over {-s}\right)
- {1 \over 4} \log \left({-s} \over {m^2_e}\right)
+ {7 \over {18}}
\right ],
\end{equation}

\noindent where $s = \left( k_{e^-} + k_{e^+}\right)^2$.
The log factor is of course expected, since the photon ``absorptive"
part comes from the expansion $\log (-s) = \log s + i\pi$.

This representation of the amplitude leads to a CP-conserving branching
ratio of

\begin{equation}
BR(K_L \rightarrow \pi^0 e^+ e^-)_{\rm CPC} = 4.89 \times 10^{-12}
\end{equation}

\noindent for $a_V = -0.96$. More generally we show the CP-conserving
branching ratio vs. $a_V$ in Fig. 10. Note that while for many values
of $a_V$ the CP-conserving rate is small compared to the CP-violating
rate of previous sections, these two rates are comparable for some
range of parameters.
Note that there is no interference in the rate between the CP-conserving and
violating components, so that the rates just add, as shown in Fig. 11.
However, there can be a CP-odd asymmetry in the electron positron
energies, which we turn to in the next section.

\begin{figure}[!htbp]
\vfill
\centerline{
\begin{minipage}[t]{.44\linewidth}\centering
\mbox{\epsfig{file=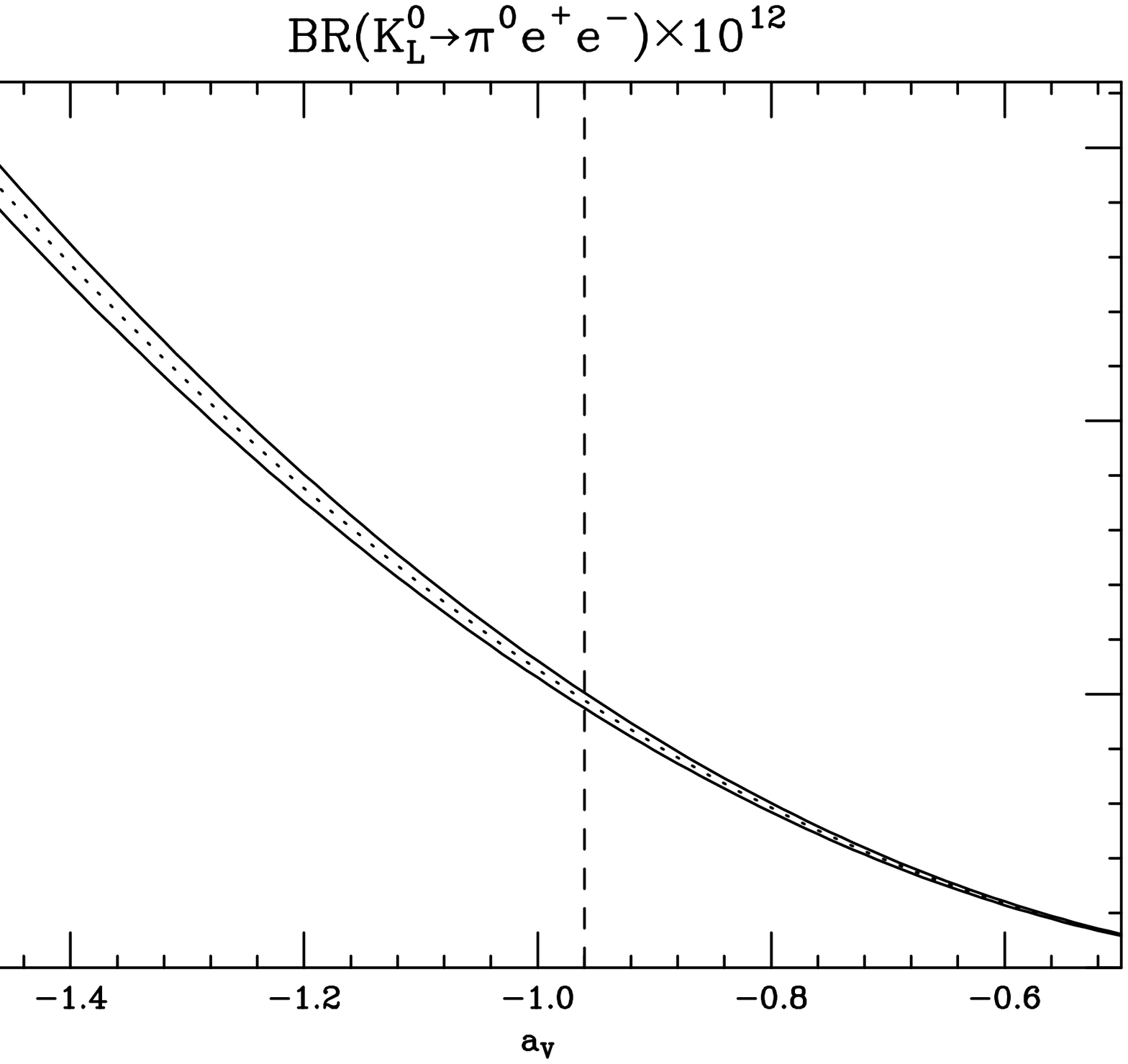,width=6.5cm}}
\caption{The CP-conserving branching ratio
$BR(K_L \ri \pi^0 e^+ e^-)_{\rm CPC}$ is plotted against $a_V$. The
convention for the solid and dashed curves is the same as for Fig. 6.
The assumed value for $a_V$ is indicated by the dashed vertical line.}
\end{minipage}
\hspace{1.0cm}
\begin{minipage}[t]{.44\linewidth}\centering
\mbox{\epsfig{file=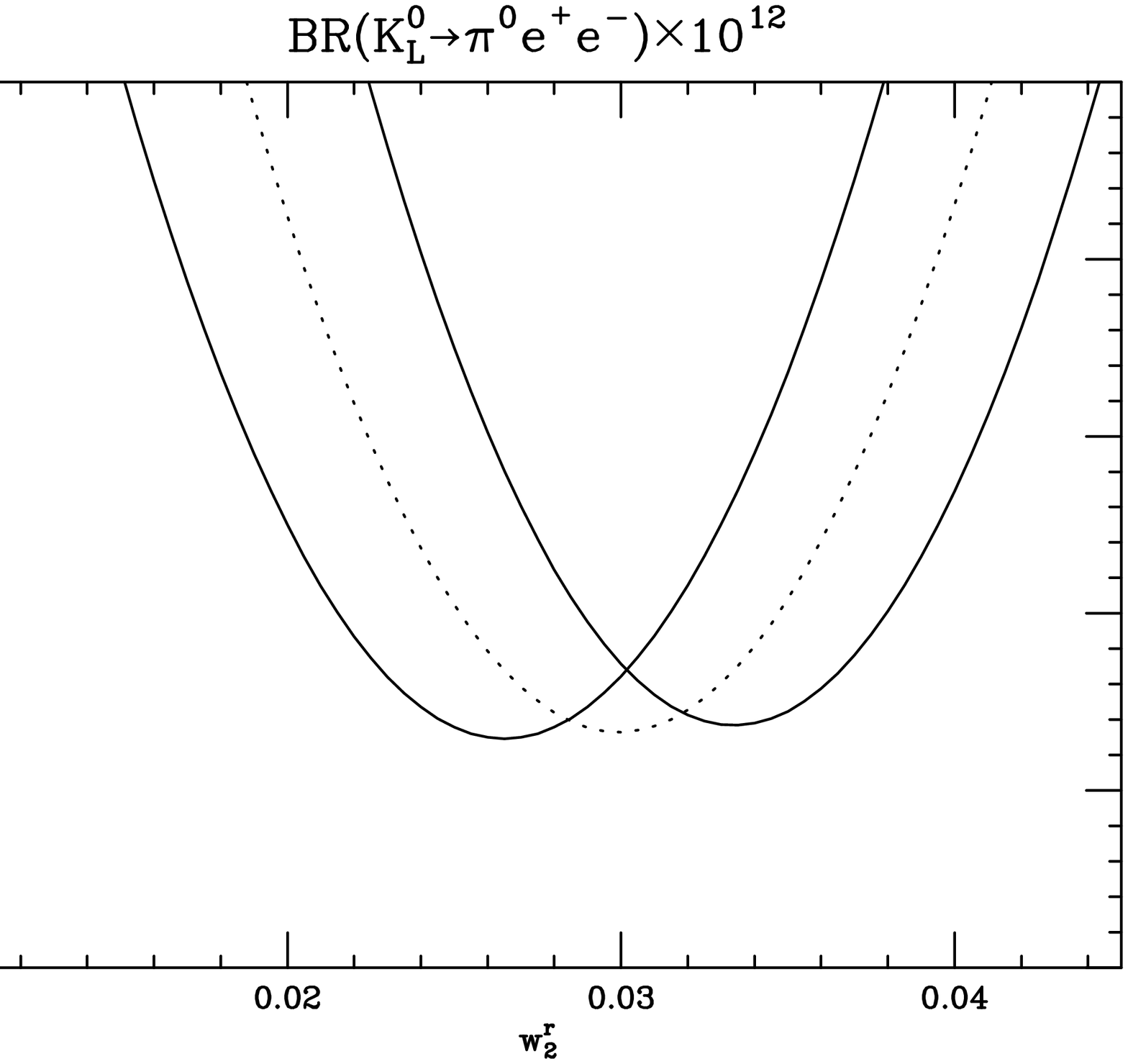,width=6.5cm}}
\caption{The complete branching ratio
$BR(K_L \ri \pi^0 e^+ e^-)$ is plotted against $w_2$. The
convention for the solid and dashed curves is the same as for Fig. 6.
$a_V = -0.96$ is assumed.}
\end{minipage}}
\end{figure}

\section{The Electron Energy Asymmetry}

When both CP-violating and CP-conserving amplitudes contribute to the
decay, there will be an asymmetry in the electron positron energy
distribution

\begin{equation}
A = {N(E_+ > E_-) - N(E_+ < E_-) \over N(E_+ > E_-) + N(E_+ < E_-)}.
\end{equation}

\noindent This will be quite large if the two amplitudes are comparable. In
contrast to the overall decay rate, this asymmetry is an unambiguous signal
of CP violation. It may be even more useful if it can be used to prove the
existence of direct CP violation. This can occur because the asymmetry is
sensitive to the phase of the CP-violating amplitude, and mass matrix CP
violation has a unique phase (that of $\epsilon$), while direct CP violation
will in general have a different phase.

For the electron energy asymmetry to be useful as a diagnostic of the form
of CP violation, the CP-conserving two-photon amplitudes must be well
known. As we discussed in the previous section, it is reasonable to expect
that this will be true in the future after further phenomenology of $K_L
\rightarrow \pi^0 \gamma \gamma$. For illustrative purposes, we will use
$a_V = -0.96$. The analysis below will need to be redone in the future if this
value of $a_V$ changes significantly. However the pattern of the analysis
and the general conclusions will be valid for a wide range of values of
$a_V$.

The amplitudes involved have been given in the previous section. Note that
the two-photon amplitude has both real and imaginary parts, the mass matrix
CP-violating amplitude has a phase of $45^o$, while the direct component
is purely imaginary. The asymmetry is proportional to the imaginary part of
$B$ times the real part of $d_V$ minus the product of the real part of
$B$ and imaginary part of $d_V$. The asymmetry may be defined in
differential form:

\begin{eqnarray}
{d \Gamma \over dz} (E_+ > E_-) & = & \int_0^{{1 \over 2} \lambda^{{1
\over 2}} (1, z, r^2)} dy {d \Gamma \over dy dz}, \nonumber \\
{d \Gamma \over dz} (E_+ < E_-) & = & \int^0_{-{1 \over 2}
\lambda^{{1 \over 2}} (1, z, r^2)} dy {d \Gamma \over dy dz}, \nonumber
\\
\lambda^{{1 \over 2}} (1, z, r^2 ) & = & 1 + z^2 + r^4 - 2z - 2r^2 - 2r^2z,
\nonumber \\
r & = & {m_{\pi} \over m_K}, \nonumber \\
A(z) & \equiv & {{d \Gamma (E_+ > E_-)/dz - d \Gamma
 (E_+ < E_-)/dz} \over {d \Gamma (E_+ > E_-) /dz + d \Gamma(E_+ <
E_-)/dz}}, \nonumber \\
z & \equiv & (k_{e^-} + k_{e^+})^2/m^2_K.
\end{eqnarray}

\noindent In Fig. 12 we plot the differential asymmetry for several values of
$w_2$ in the case when there is no direct CP violation. Fig. 13 gives the
same information for ${\rm Im} \lambda_t = 10^{-4}$. We see that the
asymmetry is sizable for many values of $w^{\phantom{2}}_2$ and that it depends
significantly on direct CP violation, even possibly changing sign. The
integrated asymmetry is plotted versus $w_2$ in Figs. 14, 15.

\begin{figure}[!htbp]
\vfill
\centerline{
\begin{minipage}[t]{.44\linewidth}\centering
\mbox{\epsfig{file=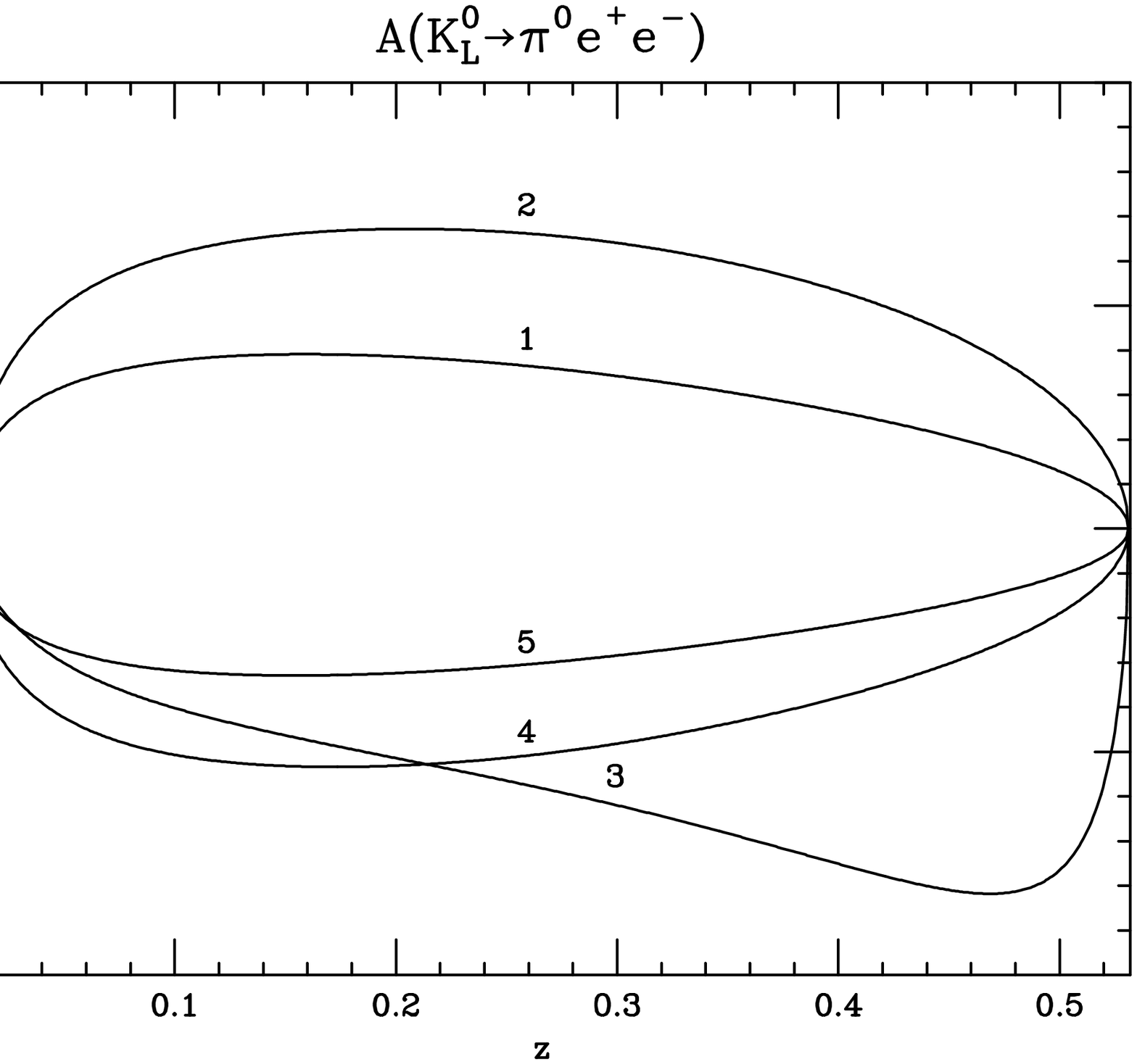,width=6.5cm}}
\caption{The differential asymmetry $A(K_L \ri \pi^0 e^+ e^-)$ is plotted
against $z \equiv (k + k^{\prime})^2/m^2_K$ for $w_2$ = 1 $\times
10^{-2}$ (curve 1), 2 $\times 10^{-2}$ (curve 2), $\dots$,
5 $\times 10^{-2}$ (curve 5), in the case when there is no direct
CP violation. $a_V = -0.96$ is assumed.}
\end{minipage}
\hspace{1.0cm}
\begin{minipage}[t]{.44\linewidth}\centering
\mbox{\epsfig{file=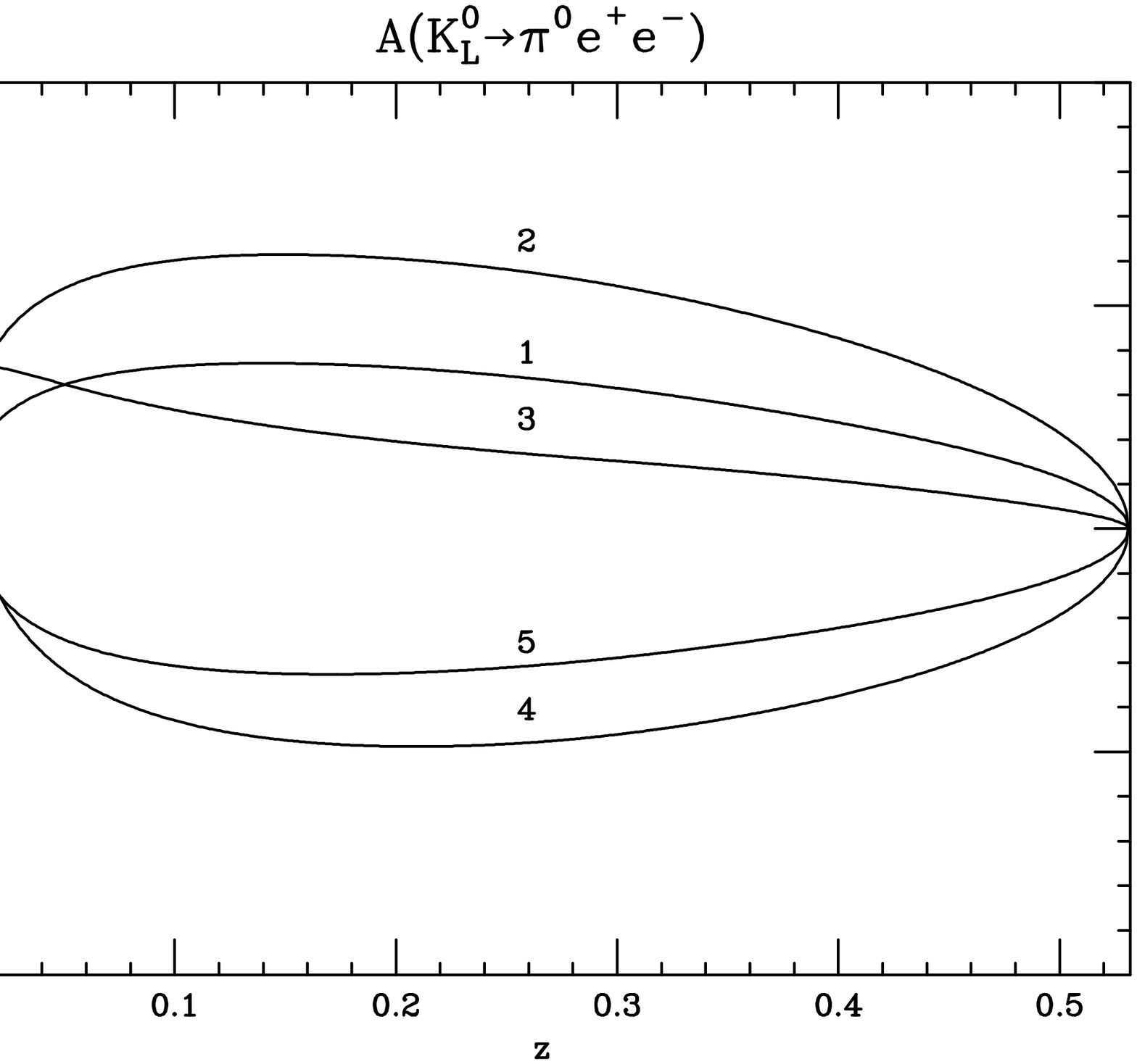,width=6.5cm}}
\caption{Same as Fig. 12 for ${\rm Im}\lambda_t$ = $10^{-4}$.}
\end{minipage}}
\end{figure}

\begin{figure}[!htbp]
\vfill
\centerline{
\begin{minipage}[t]{.44\linewidth}\centering
\mbox{\epsfig{file=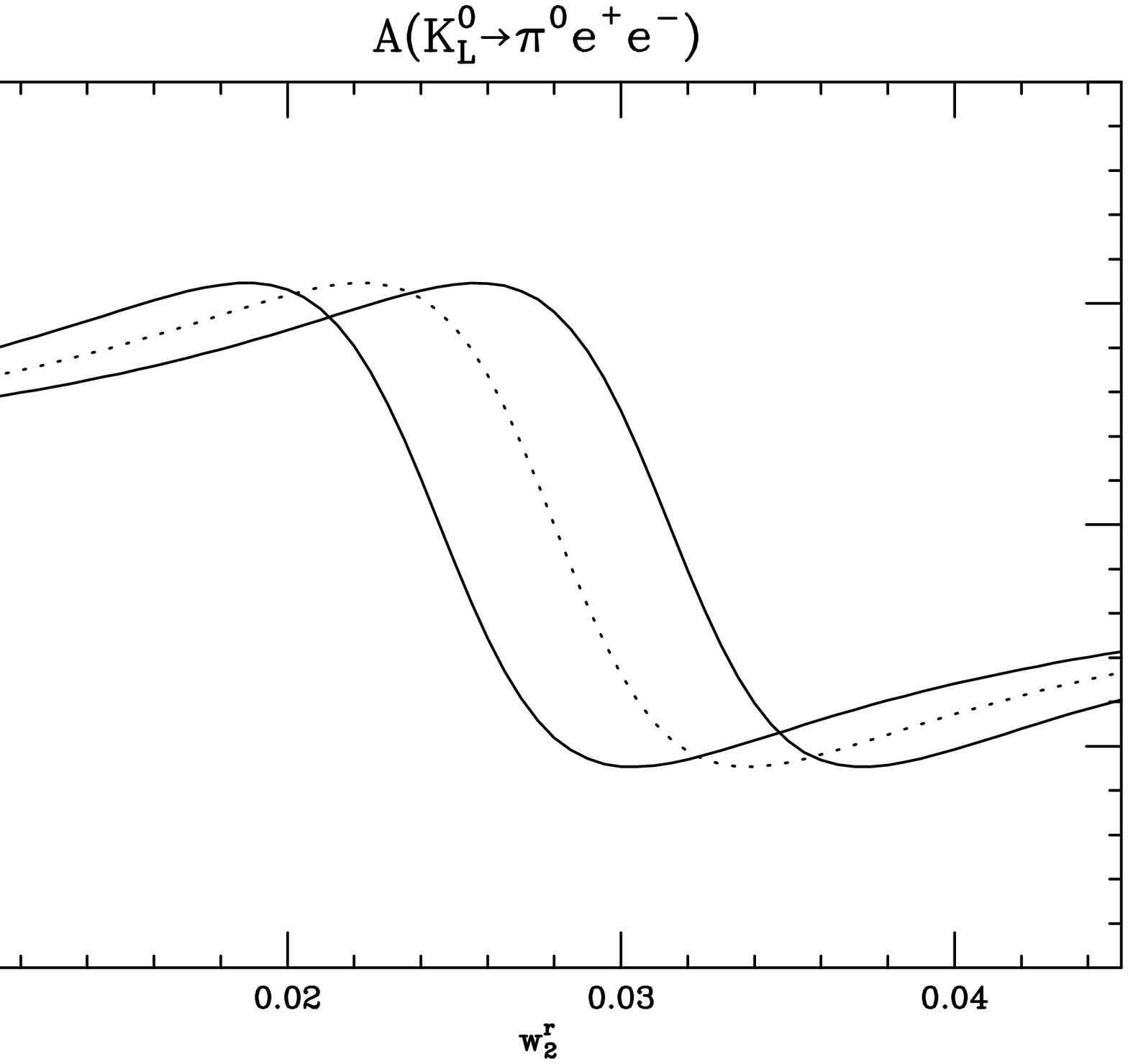,width=6.5cm}}
\caption{The integrated asymmetry is plotted vs. $w_2$ in
the case when there is no direct CP violation. The convention
for the solid and dashed curves is the same as for Fig. 6. $a_V =
-0.96$ is assumed.}
\end{minipage}
\hspace{1.0cm}
\begin{minipage}[t]{.44\linewidth}\centering
\mbox{\epsfig{file=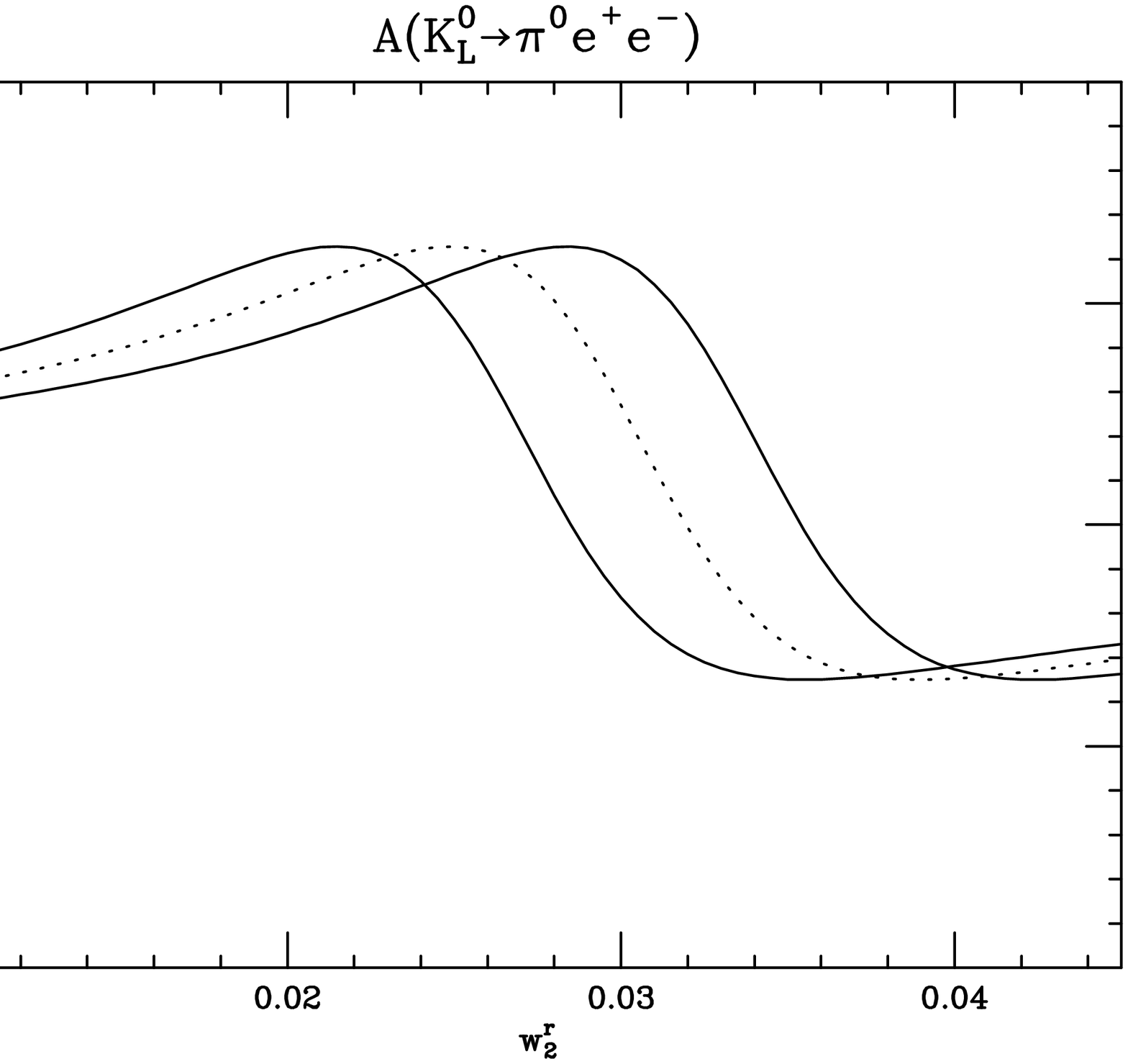,width=6.5cm}}
\caption{Same as Fig. 14 for ${\rm Im}\lambda_t$ = $10^{-4}$.}
\end{minipage}}
\end{figure}

Both the decay rate and the asymmetry depend on $w^{\phantom{2}}_S$
and this forms the main uncertainty in the analysis. However, it is
possible to remove this uncertainty by measuring both observables.
For a given value of the CP-conserving amplitude, there is a strict
correlation between the two. In Fig. 16 we plot the values of the
branching ratio and the integrated asymmetry $A$ as one varies $w_2$.
We see that the curves with and without direct CP violation are well
separated for much of the range. To the extent that we understand the
CP-conserving amplitude, this can be used as a diagnostic test for
direct CP violation.

\section{Time dependent interference of $K_{\bf 1}$, $K_{\bf 2}$${\bf \ri}$
${\bf \pi}^{\bf 0}{\bf e}^{\bf +}{\bf e}^{\bf -}$}

Littenberg \cite{L} has suggested a time dependent analysis of a state
starting out as a $K^0$ in order to extract maximal information about
the decay mechanism. This is far more demanding experimentally then
simply measuring $K_L$ or $K_S$ decays separately. However we analyze
this technique in order to assess its usefulness.

A state that at $t=0$ is a $K^0$ will evolve into a mixture of $K_L$
and $K_S$:

\begin{equation}
\vert K^0(t) \rangle = {1 \over {(1 + \epsilon)
\sqrt{2}}}\left\{e^{-iH_S t} \left[ \vert K_1 \rangle + \epsilon \vert
K_2 \rangle \right] + e^{-iH_L t}\left[\vert K_2 \rangle + \epsilon \vert
K_1 \rangle \right] \right\},
\end{equation}

\noindent where

\begin{equation}
H_j = m_j - i{\Gamma_j \over 2}, \qquad j={\rm S,L}.
\end{equation}

\noindent Ignoring small effects such as second order CP violation and
direct CP-violating components in the parameter $\epsilon$, we then
have a time development proportional to

\begin{eqnarray}
\vert \langle \pi^0 e^+ e^- \left \vert {\cal H} \right \vert K^0(t) \rangle
\vert^2 & \approx & {1 \over 2} \left\{ \left \vert A_S \right \vert^2
e^{-\Gamma_S t} +
\left \vert \epsilon A_S + A_{\rm dir} + A_{\rm CPC} \right \vert^2
e^{-\Gamma_L t} \phantom{e^{-{(\Gamma_L + \Gamma_S) t} \over 2}}
\right .\nonumber \\
& + & \left. 2 {\rm Re} \left [ \left (\epsilon A_S + A_{\rm dir} + A_{\rm
CPC} \right ) A^*_S e^{-i (m_L -m_S) t} \right ] e^{-{{(\Gamma_L +
\Gamma_S) t} \over 2}} \right \}.
\end{eqnarray}

Here $A_{\rm dir}$, $\epsilon A_S$ and $A_{\rm CPC}$ are the
amplitudes analyzed in Secs. III, IV and VI respectively. Measurements at
early time $(t \ll 1/\Gamma_S)$ determine $\Gamma(K_S \ri \pi^0 e^+
e^-)$ while at the late time $(t \gg 1/\Gamma_S)$ one observes
$\Gamma(K_L \ri \pi^0 e^+ e^-)$. These contain the information
described in preceding sections. However in the interference region
$t$ = ${\cal O} \left( {1/(m_L - m_S)} \right)$ $\sim$ ${\cal O}
(1/\tau^{\phantom{\dagger}}_S)$, we obtain extra information about
the separate contributions to the decay amplitude.

In Fig. 17 we show the time dependent signal for the cases of pure
mass matrix CP violation and the addition of direct CP violation,
using the analysis of the previous sections with $w_2 = 4 L_9$. We see
that the shape of the interference region does differentiate these two
cases, but that for the case studied the dependence on direct CP
violation is not so large as to allow an easy experimental
determination.

\begin{figure}[!htbp]
\vfill
\centerline{
\begin{minipage}[t]{.44\linewidth}\centering
\mbox{\epsfig{file=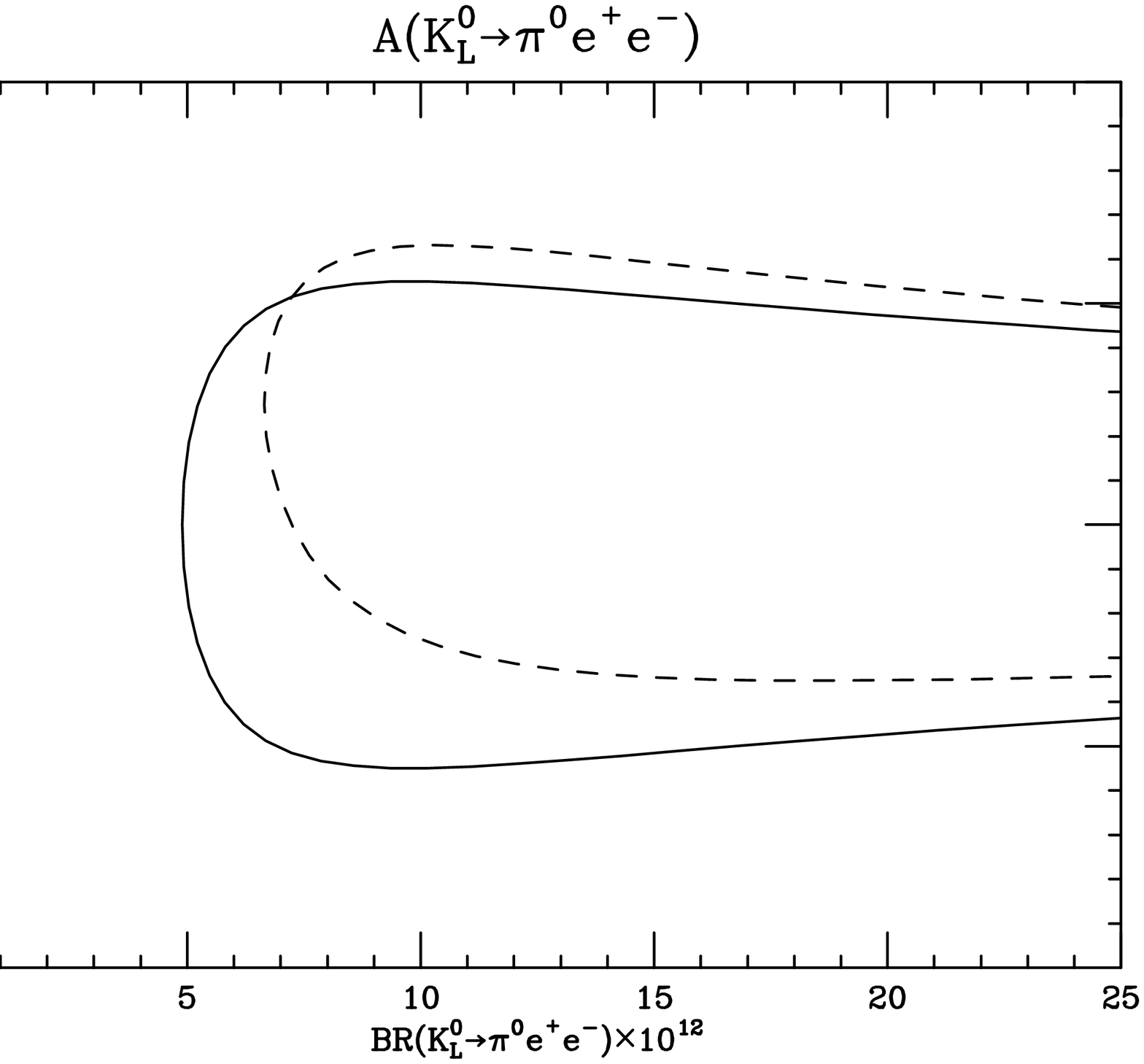,width=6.5cm}}
\caption{The integrated asymmetry A is plotted vs. the
branching ratio $BR(K_L \ri \pi^0 e^+ e^-)$ as one varies $w_2$. $a_V
= -0.96$ is assumed. The
solid curve describes the case when there is no direct CP violation,
the dashed curve is for ${\rm Im}\lambda_t$ = $10^{-4}$.}
\end{minipage}
\hspace{1.0cm}
\begin{minipage}[t]{.44\linewidth}\centering
\mbox{\epsfig{file=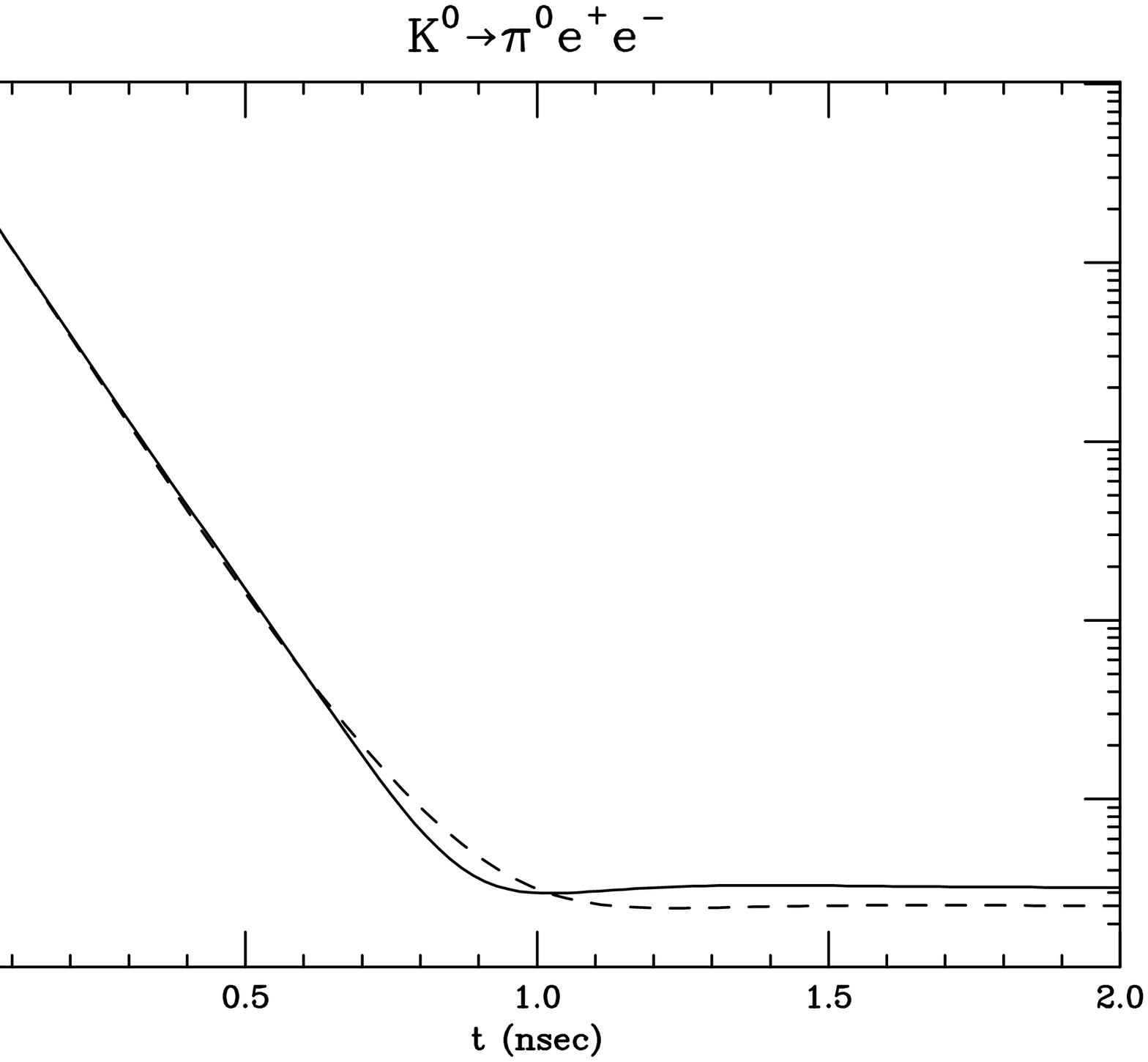,width=6.5cm}}
\caption{The normalized time distribution of $K^0 \ri \pi^0 e^+ e^-$
decays is plotted vs. time $t$ in nsecs. $a_V = -0.96$ is assumed.
The solid curve describes the case when there is no direct CP
violation, the dashed curve is for ${\rm Im}\lambda_t$ = $10^{-4}$.}
\end{minipage}}
\end{figure}

\section{Summary}

Because there are three possible contributions to $K_L \ri \pi^0 e^+
e^-$, the analysis has multiple issues which theory must address. We
have provided an updated analysis of all of the components. The goal
of identifying direct CP violation will not be easily accomplished.
The decay rate by itself suffers from a severe uncertainty in the
analysis of the mass matrix contributions. In the chiral analysis
there is a free parameter, $w_2$, which is not fixed experimentally,
and which has a strong influence on the decay rate. Measuring the
related rate of $K_S \ri \pi^0 e^+ e^-$ would determine this parameter
and will likely allow us to determine whether or not direct CP
violation is present.

Alternatively if the $K_S$ branching ratio is not measured it may be
possible to signal direct CP violation using the asymmetry in the $e^+$,
$e^-$ energies. The direct and mass matrix CP violations have
different phases, and the asymmetry is sensitive to their difference.
There exist some combinations of the parameters where there remains
some ambiguity, but for sizeable portions of the parameter space
direct CP violation can be signaled by a simultaneous measurement of
the decay rate and the energy asymmetry, as illustrated in Fig. 16.

Progress in the theoretical analysis is possible, especially in the
CP-conserving amplitude which proceeds through the two-photon
intermediate state. Here both a better phenomenological understanding
of the related decay $K_L \ri \pi^0 \gamma \gamma$, and a better
theoretical treatment of the dispersive contribution should be possible
in the near future.

\bigskip\bigskip
{\Large {\bf Acknowledgments}}
\bigskip

We would like to thank B.~R.~Holstein, J.~Kambor, L.~S.~Littenberg and
G.~Valencia for helpful discussions.

\vfill\eject

\vfill\eject

\end{document}